\documentclass[aps,prd,reprint,onecolumn,superscriptaddress,nobibnotes]{revtex4-2}
\usepackage{amsfonts,amsmath,amssymb,mathrsfs,hyperref,url}
\usepackage[mathscr]{euscript}
\usepackage{ytableau}
\numberwithin{equation}{section}

\begin{document}

%\preprint{YITP-SB-17-48}

\title{BPZ equations for higher degenerate fields and nonperturbative Dyson-Schwinger equations}

\newcommand*{\CERN}{Department of Theoretical Physics, CERN, \\ 1211 Geneva 23, Switzerland}\affiliation{\CERN}
\newcommand*{\YITP}{C.N. Yang Institute for Theoretical Physics, Stony Brook University, \\ Stony Brook, New York 11794, USA}\affiliation{\YITP}
\newcommand*{\ZJU}{Zhejiang Institute of Modern Physics, School of Physics, Zhejiang University, \\ 866 Yuhangtang Rd, Hangzhou, Zhejiang, 310058, China}\affiliation{\ZJU}
\newcommand*{\NHETC}{New High Energy Theory Center, Rutgers University, \\ Piscataway, New Jersey 08854, USA}\affiliation{\NHETC}

\author{Saebyeok Jeong}
\email{saebyeok.jeong@gmail.com}
\affiliation{\CERN}\affiliation{\YITP}\affiliation{\NHETC}

\author{Xinyu Zhang}
\email{xinyu.zhang@zju.edu.cn}
\affiliation{\ZJU}\affiliation{\YITP}\affiliation{\NHETC}

\date{\today}

\begin{abstract}
In the two-dimensional Liouville conformal field theory,  correlation functions involving a degenerate field satisfy  partial differential equations due to the decoupling of the null descendant field. 
On the other hand, the instanton partition function of a four-dimensional $\mathcal{N}=2$ supersymmetric theory in the $\Omega$-background at a special point of the parameter space also satisfies a partial differential equation resulting from the constraints of the gauge field configurations. This partial differential equation can be proved using the nonperturbative Dyson-Schwinger equations. We show for the next-to-simplest case that the partial differential equations obtained from two different perspectives can be identified, thereby confirming an assertion of the BPS/CFT correspondence.
\end{abstract}

%\pacs{11.25.Hf, 11.30.Pb} 
%11.30.Pb: Supersymmetry
%11.25.Hf: Conformal field theory

\maketitle

\section{Introduction}

Ever since the groundbreaking work of Seiberg and Witten on four-dimensional $\mathcal{N}=2$ supersymmetric gauge theories with gauge group $\mathrm{SU}(2)$ \cite{Seiberg:1994rs,Seiberg:1994aj}, there has been continuous progress in constructing and analyzing $\mathcal{N}=2$ supersymmetric theories. Although many impressive statements have already been made over the last few decades, we continue to discover new interesting results.

It has long been appreciated that when understanding a complicated system, it is helpful to explore its deformations and study the dependence on the deformation parameters. This general lesson has been convincingly demonstrated in \cite{Nekrasov:2002qd}. The four-dimensional $\mathcal{N}=2$ supersymmetric gauge theories were studied in the $\Omega$-background $\mathbb{R}_{\varepsilon{1},\varepsilon_{2}}^{4}$, with two deformation parameters $\varepsilon_{1}$ and $\varepsilon_{2}$. The deformed theory breaks Poincar\'e symmetry in a rotationally covariant way while still preserving a combination of the deformed supersymmetry. Applying the localization technique, the supersymmetric partition function $\mathcal{Z}$ and correlation functions of $\mathcal{N}=2$ chiral operators have been computed exactly for a large class of four-dimensional $\mathcal{N}=2$ supersymmetric gauge theories. At generic points in the parameter space, the low-energy effective prepotential $\mathcal{F}_{\mathrm{eff}}$ can be extracted from $\mathcal{Z}$ by taking the flat-space limit \cite{Nekrasov:2002qd,Nekrasov:2003rj},
\begin{equation}
\mathcal{F}_{\mathrm{eff}}=-\lim_{\varepsilon_{1},\varepsilon_{2}\to0}\varepsilon_{1}\varepsilon_{2}\log\mathcal{Z}.
\end{equation}
Further investigation of the $\Omega$-background with finite $\varepsilon_{1},\varepsilon_{2}$ resulted in the proposal of a remarkable relation, the BPS/CFT correspondence, which identifies correlation functions of $\mathcal{N}=2$ chiral operators with quantities in two-dimensional conformal field theories or deformations thereof.

After establishing the BPS/CFT correspondence, we expect to gain insights into four-dimensional $\mathcal{N}=2$ supersymmetric gauge theories using the knowledge of two-dimensional theories, and vice versa. One particular implementation of the BPS/CFT correspondence is the Alday-Gaiotto-Tachikawa (AGT) correspondence, which was first conjectured as a relationship between a class of superconformal $\mathrm{SU}(2)$ quiver gauge theories and the Liouville conformal field theory \cite{Alday:2009aq}, and was soon extended to a more general relationship between quiver gauge theories with other gauge groups and the Toda conformal field theory \cite{Wyllard:2009hg}.

The four-dimensional $\mathcal{N}=2$ superconformal field theories considered in the AGT correspondence can be obtained by compactifying the six-dimensional $\mathcal{N}=(2,0)$ superconformal theory on a punctured Riemann surface $\mathcal{C}$ \cite{Gaiotto:2009we}. When the theory admits a weakly-coupled Lagrangian description, we can often compute its partition function in the $\Omega$-background \cite{Nekrasov:2002qd,Nekrasov:2012xe}. It was proposed that the instanton part of the partition function $\mathcal{Z}^{\mathrm{instanton}}$ can be identified with a conformal block in the Liouville/Toda conformal field theory, and the partition function on a (squashed) sphere $S_{b}^{4}$ \cite{Pestun:2007rz,Hama:2012bg}, which is given by the integral of the absolute value squared of the full partition function, can be identified with correlation functions in the Liouville/Toda conformal field theory on $\mathcal{C}$. Based on careful analysis of the structure of the instanton moduli space, some versions of the AGT correspondence have been proved \cite{schiffmann2013cherednik,Maulik:2012wi,Bourgine:2015szm}.

In the study of the AGT correspondence, it is often assumed that the Coulomb branch parameters take generic values. However, it is also interesting to specialize these parameters and explore the consequence of the AGT correspondence in the $\mathcal{N}=2$ gauge theory context. In two-dimensional conformal field theory, we can make one of the fields in the correlation function degenerate. Belavin, Polyakov, and Zamolodchikov (BPZ) showed that such correlation functions satisfy partial differential equations as a result of the decoupling of the null descendant field \cite{Belavin:1984vu,Fateev:2009me}. The order of the differential equation is the level of the null field in the corresponding degenerate representation. In the case of Toda field theories, similar differential equations have been derived for certain four-point correlation functions in \cite{Fateev:2005gs,Fateev:2007ab}. Correspondingly, the gauge field configurations in the four-dimensional $\mathcal{N}=2$ superconformal quiver gauge theories are constrained. We will show that the corresponding instanton partition functions also satisfy partial differential equations. In the context of the $\mathcal{N}=2$ gauge theory, the specialization of the Coulomb parameter initiate partial Higgsing of the theory, producing a half-BPS surface defect as a result \cite{Dorey:2011pa,Nekrasov:2017rqy}. The differential equation we obtained is the quantized chiral ring relation of the so-obtained 2D/4D coupled system \cite{Jeong:2018qpc}. This program was investigated carefully in the Nekrasov-Shatashvili limit \cite{Nekrasov:2009rc} of the $\Omega$-background, which corresponds to the classical limit $c\to\infty$ of two-dimensional conformal field theories \cite{Mironov:2009uv,Mironov:2009dv,Maruyoshi:2010iu,Poghossian:2010pn,Marshakov:2010fx,Fucito:2013fba,Litvinov:2013sxa,Nekrasov:2013xda,Poghossian:2016rzb,Poghosyan:2016mkh}. However, previous methods become less powerful when we would like to go beyond such limits. Our results provide a new method to go beyond this limit.

In this paper, we shall follow the idea of \cite{Nekrasov:2017gzb} to provide a derivation of the differential equation using the nonperturbative Dyson-Schwinger (NPDS) equations, which result from the fact that the path integral of the instanton partition function is invariant with respect to the transformations changing topological sectors of the field space. We review the result of \cite{Nekrasov:2017gzb} and study the case of $\mathrm{U}(2)$ superconformal linear quiver gauge theories with the next-to-simplest constraint in this paper. Similar methods have also been applied to the study of Bethe/gauge correspondence \cite{Nekrasov:2009uh,Nekrasov:2009ui,Nekrasov:2009rc} in \cite{Jeong:2017pai}.

The rest of the paper is organized as follows. In Sec. \ref{sec:degenerateCFT}, we recall some basic facts about two-dimensional Liouville field theory and review the derivation of BPZ equations on the degenerate correlation functions. In Sec. \ref{sec:4dgauge}, we review the relevant details of four-dimensional $\mathcal{N}=2$ quiver gauge theories in the $\Omega$-background. We summarize the result of the partition function and review the NPDS equations. In Sec. \ref{sec:A1}, we study the superconformal gauge theory with gauge group $\mathrm{U}(N)$. We show that the instanton partition function at the simplest nontrivial degenerate point in the parameter space is a (generalized) hypergeometric function. After working out this simple warm-up example, we consider the $\mathrm{U}(2)$ superconformal linear quiver gauge theory in Sec. \ref{sec:quiver}. We review the second-order differential equation on the instanton partition function derived in \cite{Nekrasov:2017gzb} and derive the third-order differential equation for the next-to-simplest case. We also identify the differential equations derived from both sides using the AGT dictionary. Finally, we conclude in Sec. \ref{sec:Conclusions} and discuss possible directions for future work. In Appendix \ref{sec:hypergeometric}, we review some standard material on the (generalized) hypergeometric function. In Appendix \ref{sec:U1}, we derive the partition function of the $\mathrm{U}(1)$ factor using the NPDS equations.

\section{Degenerate correlation functions in the Liouville field theory \label{sec:degenerateCFT}}

In this section, we recall some basic facts about two-dimensional
Liouville field theory and present the derivation of the BPZ equations
on the degenerate correlation functions. 

\subsection{Degenerate fields in the Liouville field theory}

The two-dimensional Liouville conformal field theory is defined by the action,
\begin{equation}
S_{\mathrm{Liouville}}=\int d^{2}\sigma\sqrt{g}\left(\frac{1}{4\pi}\partial_{a}\phi\partial^{a}\phi+\mu e^{2b\phi}+\frac{Q}{4\pi}R\phi\right),
\end{equation}
where the background charge $Q=b+b^{-1}$, and $R$ is the Ricci scalar
of the Riemann surface. The symmetry algebra of the theory is two
independent copies of the Virasoro algebra, with the central charge
$c=1+6Q^{2}$. In the following, we focus on the chiral part, which
is spanned by generators $L_{n}$ for $n\in\mathbb{Z}$ and the central
charge $c$, satisfying
\begin{equation}
\left[L_{m},L_{n}\right]=\left(m-n\right)L_{m+n}+\frac{c}{12}\left(m^{3}-m\right)\delta_{m+n,0}.
\end{equation}

For the Virasoro algebra, a conformal primary field $V_{\Delta}$
with the conformal dimension $\Delta$ is defined to be 
\begin{equation}
L_{n>0}V_{\Delta}=0,\quad L_{0}V_{\Delta}=\Delta V_{\Delta}.
\end{equation}
The descendant fields are obtained by taking the linear combinations
of the basis vectors $L_{-\vec{n}}V_{\Delta}=L_{-n_{1}}L_{-n_{2}}\cdots L_{-n_{l}}V_{\Delta}$,
where $\vec{n}=\left\{ 1\leq n_{1}\leq n_{2}\leq\cdots\leq n_{l}\right\} $.
The conformal dimension of the basis vector $L_{-\vec{n}}V_{\Delta}$
is $\Delta+|\vec{n}|$, where the number $|\vec{n}|=\sum_{i=1}^{l}n_{i}$
is called the level of $L_{-\vec{n}}V_{\alpha}$. 

A primary field $V_{\Delta}$ is called degenerate if it has a null
descendant field $\tilde{V}=\sum_{\vec{n}}C_{\vec{n}}L_{-\vec{n}}V_{\Delta}\neq V_{\Delta}$,
such that $L_{n}\tilde{V}=0$ for $n>0$. If the null field is at
the level one, then 
\begin{equation}
L_{n}\left(L_{-1}V_{\Delta}\right)=0,\quad n>0.
\end{equation}
This is automatically true for $n\geq2$, and for $n=1$ we have
\begin{equation}
0=L_{1}L_{-1}V_{\Delta}=2L_{0}V_{\Delta}=2\Delta V_{\Delta}.
\end{equation}
Thus the field $V_{\Delta}=1$ with zero conformal dimension. If the
level-two descendant field $\tilde{V}=\left(C_{1,1}L_{-1}^{2}+C_{2}L_{-2}\right)V_{\Delta}$
is null, then $L_{n}\tilde{V}=0$ for $n\geq1$. The nontrivial constraints
are
\begin{align}
0 & =  L_{1}\tilde{V}=\left(\left(4\Delta+2\right)C_{1,1}+3C_{2}\right)L_{-1}V_{\Delta},\nonumber \\
0 & =  L_{2}\tilde{V}=\left(6\Delta C_{1,1}+\left(4\Delta+\frac{c}{2}\right)C_{2}\right)V_{\Delta}.
\end{align}
Therefore, we have
\begin{equation}
\begin{vmatrix}4\Delta+2 & 3\\
6\Delta & 4\Delta+\frac{c}{2}
\end{vmatrix}=0,
\end{equation}
which gives two solutions
\begin{align}
\Delta_{(2,1)} & =  -\frac{1}{2}-\frac{3}{4}b^{2},\quad\tilde{V}_{(2,1)}=\left(\frac{1}{b^{2}}L_{-1}^{2}+L_{-2}\right)V_{\Delta_{(2,1)}},\label{eq:Deg21}\\
\Delta_{(1,2)} & =  -\frac{1}{2}-\frac{3}{4b^{2}},\quad\tilde{V}_{(1,2)}=\left(b^{2}L_{-1}^{2}+L_{-2}\right)V_{\Delta_{(1,2)}}.
\end{align}
If the level-three descendant field $\tilde{V}=\left(C_{1,1,1}L_{-1}^{3}+C_{1,2}L_{-1}L_{-2}+C_{3}L_{-3}\right)V_{\Delta}$
is null, then $L_{n}\tilde{V}=0$ for $n\geq1$. Since $L_{3}=\left[L_{2},L_{1}\right]$,
we only need 
\begin{align}
0 & = L_{1}\tilde{V}=\left(\left(2\Delta+4\right)C_{1,2}+4C_{3}\right)L_{-2}V_{\Delta}+\left(\left(6\Delta+6\right)C_{1,1,1}+3C_{1,2}\right)L_{-1}^{2}V_{\Delta},\nonumber \\
0 & =  L_{2}\tilde{V}=\left(\left(6+18\Delta\right)C_{1,1,1}+\left(4\Delta+9+\frac{c}{2}\right)C_{1,2}+5C_{3}\right)L_{-1}V_{\Delta}.
\end{align}
Therefore, we have
\begin{equation}
\begin{vmatrix}0 & 2\Delta+4 & 4\\
6\Delta+6 & 3 & 0\\
6+18\Delta & 4\Delta+9+\frac{c}{2} & 5
\end{vmatrix}=0,
\end{equation}
which gives two solutions
\begin{align}
\Delta_{(3,1)} & =  -1-2b^{2},\quad\tilde{V}_{(3,1)}=\left(\frac{1}{4b^{2}}L_{-1}^{3}+L_{-1}L_{-2}+\left(b^{2}-\frac{1}{2}\right)L_{-3}\right)V_{\Delta_{(3,1)}},\label{eq:Deg31}\\
\Delta_{(1,3)} & =  -1-\frac{2}{b^{2}},\quad\tilde{V}_{(1,3)}=\left(\frac{b^{2}}{4}L_{-1}^{3}+L_{-1}L_{-2}+\left(\frac{1}{b^{2}}-\frac{1}{2}\right)L_{-3}\right)V_{\Delta_{(1,3)}}.
\end{align}
Generally, the conformal dimension of a degenerate field can be read
from the Kac determinant formula, and is given by
\begin{equation}
\Delta_{(m,n)}=\frac{Q^{2}-\left(mb+nb^{-1}\right)^{2}}{4},\quad m,n\in\mathbb{Z}^{+},
\end{equation}
with the null vector being at the level $mn$.

\subsection{BPZ equations}

Now we are ready to derive the BPZ equations on the $(r+3)$-point
correlation function of the conformal primary fields, with one of
the primary fields being degenerate. In order to relate a correlation
function involving Virasoro generators acting on a primary field with
a correlation function of purely primary fields, we use the conformal
Ward identities, which state that inserting the holomorphic energy-momentum
tensor in a correlation function of primary fields yields,
\begin{equation}
\Bigg\langle T(z)\prod_{i=-1}^{r+1}V_{\Delta_{i}}(z_{i})\Bigg\rangle=\sum_{i=-1}^{r+1}\left(\frac{1}{z-z_{i}}\frac{\partial}{\partial z_{i}}+\frac{\Delta_{i}}{(z-z_{i})^{2}}\right)\Bigg\langle\prod_{i=-1}^{r+1}V_{\Delta_{i}}(z_{i})\Bigg\rangle.
\end{equation}

The simplest nontrivial example is the second order BPZ equation.
We assume that $\Delta_{0}=\Delta_{(2,1)}$. The decoupling of the
null descendant field (\ref{eq:Deg21}) implies that the $(r+3)$-point
correlation function satisfies,
\begin{equation}
\left[\frac{1}{b^{2}}\frac{\partial^{2}}{\partial z_{0}^{2}}+\sum_{i\neq0}\left(\frac{1}{z_{0}-z_{i}}\frac{\partial}{\partial z_{i}}+\frac{\Delta_{i}}{(z_{0}-z_{i})^{2}}\right)\right]\Bigg\langle V_{\Delta_{(2,1)}}(z_{0})\prod_{i\neq0}V_{\Delta_{i}}(z_{i})\Bigg\rangle=0.\label{eq:BPZ2-1}
\end{equation}
Similarly, the third-order BPZ equation with $\Delta_{0}=\Delta_{(3,1)}$
can be derived from the decoupling of the null vector (\ref{eq:Deg31}),
\begin{align}
0 & =  \Bigg[\frac{1}{4b^{2}}\frac{\partial^{3}}{\partial z_{0}^{3}}+\frac{\partial}{\partial z_{0}}\sum_{i\neq0}\left(\frac{1}{z_{0}-z_{i}}\frac{\partial}{\partial z_{i}}+\frac{\Delta_{i}}{(z_{0}-z_{i})^{2}}\right)\nonumber \\
 &   -\left(b^{2}-\frac{1}{2}\right)\sum_{i\neq0}\left(\frac{1}{(z_{0}-z_{i})^{2}}\frac{\partial}{\partial z_{i}}+\frac{2\Delta_{i}}{(z_{0}-z_{i})^{3}}\right)\Bigg]\Bigg\langle V_{\Delta_{(3,1)}}(z_{0})\prod_{i\neq0}V_{\Delta_{i}}(z_{i})\Bigg\rangle.\label{eq:BPZ3-1}
\end{align}

There are additional constraints on the correlation functions due
to the global conformal symmetry. Using the holomorphy of the energy-momentum
tensor at infinity, $T(z)=\mathcal{O}\left(z^{-4}\right)$ as $z\to\infty$,
we deduce the global conformal Ward identities,
\begin{align}
\left[\sum_{i=-1}^{r+1}\frac{\partial}{\partial z_{i}}\right]\Bigg\langle\prod_{i=-1}^{r+1}V_{\Delta_{i}}(z_{i})\Bigg\rangle & =  0,\label{eq:Ward1}\\
\left[\sum_{i=-1}^{r+1}\left(z_{i}\frac{\partial}{\partial z_{i}}+\Delta_{i}\right)\right]\Bigg\langle\prod_{i=-1}^{r+1}V_{\Delta_{i}}(z_{i})\Bigg\rangle & =  0,\label{eq:Ward2}\\
\left[\sum_{i=-1}^{r+1}\left(z_{i}^{2}\frac{\partial}{\partial z_{i}}+2z_{i}\Delta_{i}\right)\right]\Bigg\langle\prod_{i=-1}^{r+1}V_{\Delta_{i}}(z_{i})\Bigg\rangle & =  0.\label{eq:Ward3}
\end{align}
For our purpose, it is convenient to get rid of all the $\partial_{-1}$
and $\partial_{r+1}$ terms using (\ref{eq:Ward1}) and (\ref{eq:Ward3}),
\begin{align}
\frac{\partial}{\partial z_{-1}}\Bigg\langle\prod_{i=-1}^{r+1}V_{\alpha_{i}}(z_{i})\Bigg\rangle & =  \left[\sum_{i=0}^{r}\frac{z_{i}^{2}-z_{r+1}^{2}}{z_{r+1}^{2}-z_{-1}^{2}}\frac{\partial}{\partial z_{i}}+\sum_{i=-1}^{r+1}\frac{2z_{i}\Delta_{i}}{z_{r+1}^{2}-z_{-1}^{2}}\right]\Bigg\langle\prod_{i=-1}^{r+1}V_{\Delta_{i}}(z_{i})\Bigg\rangle,\nonumber \\
\frac{\partial}{\partial z_{r+1}}\Bigg\langle\prod_{i=-1}^{r+1}V_{\alpha_{i}}(z_{i})\Bigg\rangle & =  \left[\sum_{i=0}^{r}\frac{z_{i}^{2}-z_{-1}^{2}}{z_{-1}^{2}-z_{r+1}^{2}}\frac{\partial}{\partial z_{i}}+\sum_{i=-1}^{r+1}\frac{2z_{i}\Delta_{i}}{z_{-1}^{2}-z_{r+1}^{2}}\right]\Bigg\langle\prod_{i=-1}^{r+1}V_{\Delta_{i}}(z_{i})\Bigg\rangle.\label{eq:Ward13-2}
\end{align}
We then fix $z_{-1}=\infty$ and $z_{r+1}=0$, and the remaining global
conformal Ward identity (\ref{eq:Ward2}) gives 
\begin{equation}
\left[\sum_{i=0}^{r}\left(\nabla_{i}+\Delta_{i}\right)-\Delta_{-1}+\Delta_{r+1}\right]\Bigg\langle V_{\Delta_{-1}}(\infty)\prod_{i=0}^{r}V_{\Delta_{i}}(z_{i})V_{\Delta_{r+1}}(0)\Bigg\rangle=0.\label{eq:Ward2-2}
\end{equation}
Let us decouple a prefactor from the correlation function
\begin{equation}
\Bigg\langle V_{\Delta_{-1}}(\infty)V_{\Delta_{(m,n)}}(z_{0})\prod_{i=1}^{r}V_{\Delta_{i}}(z_{i})V_{\Delta_{r+1}}(0)\Bigg\rangle=\left[\left(\prod_{i=0}^{r}z_{i}^{L_{i}}\right)\prod_{0\leq i<j\leq r}\left(1-\frac{z_{j}}{z_{i}}\right)^{T_{ij}}\right]\chi_{r+3}^{(m,n)}(\boldsymbol{z}),
\end{equation}
where $\chi_{r+3}^{(m,n)}(\boldsymbol{z})$ only depends on the ratios
of $z_{i}$, $i=0,\cdots,r$. The identity (\ref{eq:Ward2-2}) is
satisfied if 
\begin{equation}
\sum_{i=0}^{r}\left(L_{i}+\Delta_{i}\right)-\Delta_{-1}+\Delta_{r+1}=0.\label{eq:Ward2-3}
\end{equation}
Using
\begin{align}
\left[z_{i}\frac{\partial}{\partial z_{i}},\left(\prod_{i=0}^{r}z_{i}^{L_{i}}\right)\prod_{0\leq i<j\leq r}\left(1-\frac{z_{j}}{z_{i}}\right)^{T_{ij}}\right] & =  \left(L_{i}+\sum_{j=i+1}^{r}T_{ij}\frac{z_{j}}{z_{i}-z_{j}}+\sum_{j=0}^{i-1}T_{ji}\frac{z_{i}}{z_{i}-z_{j}}\right)\nonumber \\
 &   \times\left(\prod_{i=0}^{r}z_{i}^{L_{i}}\right)\prod_{0\leq i<j\leq r}\left(1-\frac{z_{j}}{z_{i}}\right)^{T_{ij}},
\end{align}
the second-order BPZ equation (\ref{eq:BPZ2-1}) can be express in
terms of $\chi_{r+3}^{(m,n)}(\boldsymbol{z})$ as
\begin{align}
0 & =  \Bigg[\frac{1}{b^{2}}\left(\nabla_{0}+L_{0}+\sum_{j=1}^{r}T_{0j}\frac{z_{j}}{z_{0}-z_{j}}\right)^{2}-\left(1+\frac{1}{b^{2}}\right)\left(\nabla_{0}+L_{0}+\sum_{j=1}^{r}T_{0j}\frac{z_{j}}{z_{0}-z_{j}}\right)\nonumber \\
 &   +\sum_{i=1}^{r}\frac{z_{0}}{z_{0}-z_{i}}\left(\nabla_{i}+L_{i}+\sum_{j=i+1}^{r}T_{ij}\frac{z_{j}}{z_{i}-z_{j}}+\sum_{j=0}^{i-1}T_{ji}\frac{z_{i}}{z_{i}-z_{j}}\right)\nonumber \\
 &   +\sum_{i=1}^{r}\frac{z_{0}^{2}\Delta_{i}}{\left(z_{0}-z_{i}\right)^{2}}+\Delta_{r+1}\Bigg]\chi_{r+3}^{(2,1)}(\boldsymbol{z}), \label{eq:bpz2}
\end{align}
and the third-order BPZ equation (\ref{eq:BPZ3-1}) becomes
\begin{align}
0 & =  \Bigg[\frac{1}{4b^{2}}\left(\nabla_{0}+L_{0}+\sum_{j=1}^{r}T_{0j}\frac{z_{j}}{z_{0}-z_{j}}\right)^{3}-\left(\frac{3}{4b^{2}}+1\right)\left(\nabla_{0}+L_{0}+\sum_{j=1}^{r}T_{0j}\frac{z_{j}}{z_{0}-z_{j}}\right)^{2}\nonumber \\
 &   +\left(\frac{1}{b^{2}}+b^{2}+\frac{3}{2}+\sum_{i=1}^{r}\frac{z_{0}^{2}\Delta_{i}}{\left(z_{0}-z_{i}\right)^{2}}+\Delta_{r+1}\right)\left(\nabla_{0}+L_{0}+\sum_{j=1}^{r}T_{0j}\frac{z_{j}}{z_{0}-z_{j}}\right)\nonumber \\
 &  +\sum_{i=1}^{r}\frac{z_{0}}{z_{0}-z_{i}}\left(\nabla_{0}+L_{0}+\sum_{j=1}^{r}T_{0j}\frac{z_{j}}{z_{0}-z_{j}}\right)\left(\nabla_{i}+L_{i}+\sum_{j=i+1}^{r}T_{ij}\frac{z_{j}}{z_{i}-z_{j}}+\sum_{j=0}^{i-1}T_{ji}\frac{z_{i}}{z_{i}-z_{j}}\right)\nonumber \\
 &   -\left(b^{2}+\frac{1}{2}\right)\sum_{i=1}^{r}\frac{z_{0}\left(2z_{0}-z_{i}\right)}{\left(z_{0}-z_{i}\right)^{2}}\left(\nabla_{i}+L_{i}+\sum_{j=i+1}^{r}T_{ij}\frac{z_{j}}{z_{i}-z_{j}}+\sum_{j=0}^{i-1}T_{ji}\frac{z_{i}}{z_{i}-z_{j}}\right)\nonumber \\
 &   -\left(2b^{2}+1\right)\left(\sum_{i=1}^{r}\frac{z_{0}^{3}\Delta_{i}}{\left(z_{0}-z_{i}\right)^{3}}+\Delta_{r+1}\right)\Bigg]\chi_{r+3}^{(3,1)}(\boldsymbol{z}), \label{eq:bpz3}
\end{align}
where we denote
\begin{equation}
\nabla_{i}=z_{i}\frac{\partial}{\partial z_{i}}.
\end{equation}
We should determine $L_{i}$ and $T_{ij}$ when we identify the BPZ
equations with the differential equations derived in the corresponding
gauge theories.

\section{Four-dimensional $\mathcal{N}=2$ quiver gauge theory in the $\Omega$-background}\label{sec:4dgauge}

In this section, we review some useful results of four-dimensional
$\mathcal{N}=2$ quiver gauge theories in the $\Omega$-background.
A detailed discussion can be found in \cite{Nekrasov:2012xe,Nekrasov:2013xda,Nekrasov:2015wsu}. 

\subsection{Partition function}

Let us consider four-dimensional $\mathcal{N}=2$ superconformal linear
quiver gauge theories with gauge group 
\begin{equation}
G=\mathrm{U}(N_{1})\times \mathrm{U}(N_{2})\times\cdots\times \mathrm{U}(N_{r}),
\end{equation}
where $N_{1}=N_{2}=\cdots=N_{r}=N$. The vector superfield splits
into a collection of vector multiplets for each gauge factor $\mathrm{U}(N_{i})$.
The matter superfields consist of $r-1$ hypermultiplets transforming
in the bifundamental representations $\left(\bar{\boldsymbol{N}}_{i},\boldsymbol{N}_{i+1}\right)$,
$N$ hypermultiplet transforming in the antifundamental representation
of $\mathrm{U}(N_{1})$, and $N$ hypermultiplet transforming in the fundamental
representation of $\mathrm{U}(N_{r})$. Here we denote by $\boldsymbol{N}_{i}$
the representation of $G$ in which the $i$th factor acts in the
defining $N$-dimensional representation, while all other factors
act trivially. It is convenient to extend the quiver by including
two frozen nodes $\mathrm{U}(N_{0})$ and $\mathrm{U}(N_{r+1})$ corresponding to the
flavor symmetry $\mathrm{U}(N)\times \mathrm{U}(N)$. The Yang-Mills coupling constant
$g_{i}$ and the theta-angle $\vartheta_{i}$ for the gauge factor
$\mathrm{U}(N_{i})$ are combined into the complexified gauge couplings 
\begin{equation}
\tau_{i}=\frac{\vartheta_{i}}{2\pi}+\mathrm{i}\frac{4\pi}{g_{i}^{2}},\quad i=1,\cdots,r.
\end{equation}
We introduce
\begin{equation}
q_{i}=\exp\left(2\pi\mathrm{i}\tau_{i}\right)=\frac{z_{i}}{z_{i-1}},\quad i=1,\cdots,r,
\end{equation}
and $q_{0}=q_{r+1}=0$. Therefore, $z_{-1}=\infty$, $z_{r+1}=0$,
and $z_{0},z_{1},\cdots,z_{r}$ are defined up to an overall rescaling.
The vacuum expectation values of the adjoint scalars in the vector
multiplet for the gauge factor $\mathrm{U}(N_{i})$ are $a_{i}=\mathrm{diag}\left(a_{i,1},\cdots,a_{i,N}\right)$.
We also encode the masses of the (anti)fundamental hypermultiplets
in $a_{0}=\mathrm{diag}\left(a_{0,1},\cdots,a_{0,N}\right)$ and $a_{r+1}=\mathrm{diag}\left(a_{r+1,1},\cdots,a_{r+1,N}\right)$.
We denote the collection of Coulomb parameters as
\begin{equation}
\boldsymbol{a}=\left\{ a_{i,\alpha}|i=0,\cdots,r+1,\alpha=1,\cdots,N\right\} .
\end{equation}

The partition function of the theory in the $\Omega$-background $\mathbb{R}_{\varepsilon_{1},\varepsilon_{2}}^{4}$
is given by a product of the classical, the one-loop, and the instanton
contributions,
\begin{equation}
\mathcal{Z}\left(\boldsymbol{a};\boldsymbol{q};\varepsilon_{1},\varepsilon_{2}\right)=\mathcal{Z}^{\mathrm{classical}}\left(\boldsymbol{a};\boldsymbol{q};\varepsilon_{1},\varepsilon_{2}\right)\mathcal{Z}^{\mathrm{1-loop}}\left(\boldsymbol{a};\varepsilon_{1},\varepsilon_{2}\right)\mathcal{Z}^{\mathrm{instanton}}\left(\boldsymbol{a};\boldsymbol{q};\varepsilon_{1},\varepsilon_{2}\right),\label{eq:fullZ}
\end{equation}
where we denote the collection of Coulomb parameters as $\boldsymbol{a}=\left\{ a_{i,\alpha}|i=0,\cdots,r+1,\alpha=1,\cdots,N\right\} $,
and the collection of coupling constants as $\boldsymbol{q}=\left\{ q_{i}|i=1,\cdots,r\right\} $.
The classical part of the partition function is simply
\begin{equation}
\mathcal{Z}^{\mathrm{classical}}\left(\boldsymbol{a};\boldsymbol{q};\varepsilon_{1},\varepsilon_{2}\right)=\prod_{i=1}^{r}\prod_{\alpha=1}^{N}q_{i}^{-\frac{1}{2\varepsilon_{1}\varepsilon_{2}}a_{i,\alpha}^{2}},
\end{equation}
and the one-loop contribution to the partition function is 
\begin{equation}
\mathcal{Z}^{1-\mathrm{loop}}\left(\boldsymbol{a};\varepsilon_{1},\varepsilon_{2}\right)=\frac{\prod_{i=0}^{r}\prod_{\alpha,\beta}\Gamma_{2}\left(a_{i,\alpha}-a_{i+1,\beta}+\varepsilon|\varepsilon_{1},\varepsilon_{2}\right)}{\prod_{i=1}^{r}\prod_{\alpha<\beta}\Gamma_{2}\left(a_{i,\alpha}-a_{i,\beta}+\varepsilon_{1}|\varepsilon_{1},\varepsilon_{2}\right)\Gamma_{2}\left(a_{i,\alpha}-a_{i,\beta}+\varepsilon_{2}|\varepsilon_{1},\varepsilon_{2}\right)},
\end{equation}
where $\varepsilon=\varepsilon_{1}+\varepsilon_{2}$, and the Barnes
double Gamma function $\Gamma_{2}\left(x|\varepsilon_{1},\varepsilon_{2}\right)$
is defined by 
\begin{equation}
\Gamma_{2}\left(x|\varepsilon_{1},\varepsilon_{2}\right)=\exp\left\{ \frac{d}{ds}|_{s=0}\left[\frac{1}{\Gamma(s)}\int_{0}^{\infty}\frac{dt}{t}t^{s}\frac{e^{-tx}}{\left(1-e^{-\varepsilon_{1}t}\right)\left(1-e^{-\varepsilon_{2}t}\right)}\right]\right\} .
\end{equation}

The instanton part of the partition function is an equivariant integration
on the instanton moduli space with respect to the maximal torus of
the gauge group and the $SO(4)$ rotation. Applying the equivariant
localization theorem, it is given by the fixed point formula as
\begin{equation}
\mathcal{Z}^{\mathrm{instanton}}\left(\boldsymbol{a};\boldsymbol{q};\varepsilon_{1},\varepsilon_{2}\right)=\sum_{\boldsymbol{Y}}\mathcal{Q}_{\boldsymbol{Y}}\mathcal{Z}^{\mathrm{instanton}}\left(\boldsymbol{a};\boldsymbol{Y};\varepsilon_{1},\varepsilon_{2}\right).\label{eq:Zinst}
\end{equation}
The sum is over all fixed points of instanton configurations, which
are labeled by the collections of Young diagrams,
\begin{equation}
\boldsymbol{Y}=\left\{ Y^{(i,\alpha)}|0\leq i\leq r+1,\alpha=1,\cdots,N\right\} ,
\end{equation}
where $Y^{(0,\alpha)}=Y^{(r+1,\alpha)}=\emptyset$. Each Young diagram
$Y$ is a finite collection of boxes $\square=(u,v)\in Y$ arranged
in left-justified rows, with the row lengths in nonincreasing order.
The total number of boxes in the Young diagram $Y$ is denoted by
$|Y|$, and the number of boxes in each row gives a partition of $|Y|$,
\begin{equation}
Y=\left(Y_{1}\geq Y_{2}\geq\cdots\geq Y_{\ell(Y)}>Y_{\ell(Y)+1}=\cdots=0\right).
\end{equation}
We define the arm-length and leg-length as
\begin{equation}
A_{\square}^{Y}=Y_{u}-v,\quad L_{\square}^{Y}=Y_{v}^{T}-u.
\end{equation}
We also define the content $c_{\square}$ of a box $\square=(u,v)\in Y$
to be
\begin{equation}
c_{\square}=\varepsilon_{1}\left(u-1\right)+\varepsilon_{2}\left(v-1\right).
\end{equation}

The weight $\mathcal{Q}_{\boldsymbol{Y}}$ is given by
\begin{equation}
\mathcal{Q}_{\boldsymbol{Y}}=\prod_{i=1}^{r}q_{i}^{k_{i}}=\prod_{i=0}^{r}z_{i}^{k_{i}-k_{i+1}},
\end{equation}
where $k_{i}$ is the instanton charge associated with the gauge factor
$\mathrm{U}(N_{i})$, and 
\begin{equation}
k_{i}=\sum_{\alpha=1}^{N}|Y^{(i,\alpha)}|.
\end{equation}
The measure $\mathcal{Z}^{\mathrm{instanton}}\left(\boldsymbol{a};\boldsymbol{Y};\varepsilon_{1},\varepsilon_{2}\right)$
is the product of factors corresponding to the field content of the
theory,
\begin{align}
\mathcal{Z}^{\mathrm{instanton}}\left(\boldsymbol{a};\boldsymbol{Y};\varepsilon_{1},\varepsilon_{2}\right) & =  \left[\prod_{i=1}^{r}\mathcal{Z}_{\mathrm{vector}}^{\mathrm{instanton}}\left(\vec{a}_{i},\vec{Y}^{(i)};\varepsilon_{1},\varepsilon_{2}\right)\right]\nonumber \\
 &   \times\left[\prod_{i=0}^{r}\mathcal{Z}_{\mathrm{bifund}}^{\mathrm{instanton}}\left(\vec{a}_{i},\vec{Y}^{(i)},\vec{a}_{i+1},\vec{Y}^{(i+1)};\varepsilon_{1},\varepsilon_{2}\right)\right],
\end{align}
where we denote $\vec{a}_{i}=\left\{ a_{i,\alpha}|\alpha=1,\cdots,N\right\} $
and $\vec{Y}^{(i)}=\left\{ Y^{(i,\alpha)}|\alpha=1,\cdots,N\right\} $.
The contribution from a bifundamental hypermultiplet is 
\begin{align}
\mathcal{Z}_{\mathrm{bifund}}^{\mathrm{instanton}}\left(\vec{a},\vec{Y},\vec{b},\vec{W};\varepsilon_{1},\varepsilon_{2}\right) & =  \prod_{\alpha,\beta=1}^{N}\prod_{\square\in Y^{(\alpha)}}\left[a_{\alpha}-b_{\beta}-\varepsilon_{1}L_{\square}^{W^{(\beta)}}+\varepsilon_{2}\left(A_{\square}^{Y^{(\alpha)}}+1\right)\right]\nonumber \\
 &   \times\prod_{\square\in W^{(\beta)}}\left[a_{\alpha}-b_{\beta}+\varepsilon_{1}\left(L_{\square}^{Y^{(\alpha)}}+1\right)-\varepsilon_{2}A_{\square}^{W^{(\beta)}}\right].
\end{align}
If we set one collection of Young diagrams to be empty, we obtain
the contributions from an (anti)fundamental hypermultiplet,
\begin{align}
\mathcal{Z}_{\mathrm{bifund}}^{\mathrm{instanton}}\left(\vec{a}_{0},\emptyset,\vec{a}_{1},\vec{Y}^{(1)};\varepsilon_{1},\varepsilon_{2}\right) & =  \prod_{\alpha,\beta=1}^{N}\prod_{\square\in Y^{(1,\beta)}}\left(a_{0,\alpha}-a_{1,\beta}-c_{\square}\right),\label{eq:antifund}\\
\mathcal{Z}_{\mathrm{bifund}}^{\mathrm{instanton}}\left(\vec{a}_{r},\vec{Y}^{(r)},\vec{a}_{r+1},\emptyset;\varepsilon_{1},\varepsilon_{2}\right) & =  \prod_{\alpha,\beta=1}^{N}\prod_{\square\in Y^{(r,\alpha)}}\left(a_{r,\alpha}-a_{r+1,\beta}+c_{\square}+\varepsilon\right),\label{eq:fund}
\end{align}
where $\varepsilon=\varepsilon_1+ \varepsilon_2$. 
The contribution of the vector multiplet can be written in terms of
that of the bifundamental hypermultiplet as
\begin{equation}
\mathcal{Z}_{\mathrm{vector}}^{\mathrm{instanton}}\left(\vec{a},\vec{Y};\varepsilon_{1},\varepsilon_{2}\right)=\mathcal{Z}_{\mathrm{bifund}}^{\mathrm{instanton}}\left(\vec{a},\vec{Y},\vec{a},\vec{Y};\varepsilon_{1},\varepsilon_{2}\right)^{-1}.
\end{equation}

\subsection{$\mathscr{Y}$-observables, qq-characters and NPDS equations}

With the fixed point formula of instanton partition function (\ref{eq:Zinst}),
we can define the expectation value of certain BPS observables $\mathcal{O}$
in the $\Omega$-background as
\begin{equation}
\langle\mathcal{O}\rangle=\mathcal{Z}^{\mathrm{instanton}}\left(\boldsymbol{a};\boldsymbol{Y};\varepsilon_{1},\varepsilon_{2}\right)^{-1}\sum_{\boldsymbol{Y}}\mathcal{Q}_{\boldsymbol{Y}}\mathcal{Z}^{\mathrm{instanton}}\left(\boldsymbol{a};\boldsymbol{Y};\varepsilon_{1},\varepsilon_{2}\right)\mathcal{O}[\boldsymbol{Y}],\label{eq:vevO}
\end{equation}
where $\mathcal{O}[\boldsymbol{Y}]$ is the value of $\mathcal{O}$
evaluated at the instanton configuration labeled by $\boldsymbol{Y}$.

The standard local observables in the $\mathcal{N}=2$ theory on $\mathbb{R}^{4}$
are gauge-invariant polynomials of the scalar components of the vector
multiplet, $\mathrm{Tr}\Phi_{i}^{n}(\boldsymbol{x})$. However, Poincar\'e
symmetry is broken in the $\Omega$-background, and the operators
$\mathrm{Tr}\Phi_{i}^{n}(\boldsymbol{x})$ are invariant under the
deformed supersymmetry of the $\Omega$-background only at $\boldsymbol{x}=0$,
the fixed point of the rotation. The $\mathscr{Y}$-observable is
constructed as the generating function of such operators,
\begin{equation}
\mathscr{Y}_{i}(x)=x^{N}\exp\left(-\sum_{n=1}^{\infty}\frac{1}{nx^{n}}\mathrm{Tr}\Phi_{i}^{n}(0)\right),\quad i=1,\cdots,r.\label{eq:DefY}
\end{equation}
Classically, $\mathscr{Y}_{i}(x)$ is equal to the characteristic
polynomial of the scalar component in the vector multiplet of the
$i$th factor of the gauge group
\begin{equation}
\mathscr{Y}_{i}(x)^{\mathrm{classical}}=\prod_{\alpha=1}^{N}\left(x-a_{i,\alpha}\right)=\det\left(x-\Phi_{i}(0)\right).
\end{equation}
Quantum mechanically, the $\mathscr{Y}$-observables receive corrections
due to the mixing between the adjoint scalar and gluinos. The value
of $\mathscr{Y}_{i}(x)$ evaluated at the instanton configuration
labeled by $\boldsymbol{Y}$ is given by
\begin{equation}
\mathscr{Y}_{i}(x)[\boldsymbol{Y}]=\prod_{\alpha=1}^{N}\left[\left(x-a_{i,\alpha}\right)\prod_{\square\in Y^{(i,\alpha)}}\frac{\left(x-\widehat{c_{\square}}-\varepsilon_{1}\right)\left(x-\widehat{c_{\square}}-\varepsilon_{2}\right)}{\left(x-\widehat{c_{\square}}\right)\left(x-\widehat{c_{\square}}-\varepsilon\right)}\right],\label{eq:Yi}
\end{equation}
where $\widehat{c_{\square}}=a_{i,\alpha}+c_{\square}$ is the shifted
content of the box $\square\in Y^{(i,\alpha)}$. For $i=0$ and $i=r+1$,
we define
\begin{equation}
\mathscr{Y}_{0}(x)=\prod_{\alpha=1}^{N}\left(x-a_{0,\alpha}\right),\quad\mathscr{Y}_{r+1}(x)=\prod_{\alpha=1}^{N}\left(x-a_{r+1,\alpha}\right).
\end{equation}

From the $\mathscr{Y}$-observables, we can build an important class
of gauge-invariant composite operators, the so-called qq-characters.
Let us denote 
\begin{equation}
\Xi_{i}(x)=\frac{\mathscr{Y}{}_{i+1}(x+\varepsilon)}{\mathscr{Y}{}_{i}(x)}.
\end{equation}
The $\ell$th fundamental qq-characters $\mathscr{X}_{\ell}(x)$ in
linear quiver gauge theories can be written as 
\begin{equation}
\mathscr{X}_{\ell}(x)=\frac{\mathscr{Y}_{0}\left(x+\varepsilon\left(1-\ell\right)\right)}{z_{0}z_{1}\cdots z_{\ell-1}}\sum_{\substack{I\subset[0,r]\\
|I|=\ell
}
}\prod_{i\in I}\left[z_{i}\Xi_{i}\left(x+\varepsilon\left(h_{I}(i)+1-\ell\right)\right)\right],\quad\ell=0,1,\cdots,r+1,\label{eq:qq-ch}
\end{equation}
where $[0,r]=\left\{ 0,1,2,\cdots,r\right\} $, and $h_{I}(i)$ is
the number of elements in $I$ which is less than $i$. As demonstrated
in \cite{Nekrasov:2015wsu}, although the qq-characters $\mathscr{X}_{\ell}(x)$
has singularities in finite $x$, its expectation value $\langle\mathscr{X}_{\ell}(x)\rangle$
is a polynomial in $x$ of degree $N$,
\begin{equation}
\langle\mathscr{X}_{\ell}(x)\rangle=T_{\ell}(x).\label{eq:npDS}
\end{equation}
These equations are called NPDS equations, and contain nontrivial
information of the instanton partition function of the theory. In
particular, the $x^{-n}$ coefficient $\mathscr{X}_{\ell}^{(-n)}$
of the large-$x$ expansion of $\mathscr{X}_{\ell}(x)$ has zero expectation
value when $n$ is a positive integer,
\begin{equation}
\langle\mathscr{X}_{\ell}^{(-n)}\rangle=0,\quad n\geq1.
\end{equation}

\subsection{Dictionary of AGT correspondence}

It is useful to summarize the dictionary of AGT correspondence in
order to make the paper self-contained. The main statement of the
AGT correspondence is an identification between the $(r+3)$-point
correlation function in the Liouville field theory with the partition
function of superconformal quiver gauge theory with gauge group $\mathrm{SU}(2)^{r}$.

Let us decompose the $\mathrm{U}(2)$ gauge group into the $\mathrm{U}(1)$ part and
the $\mathrm{SU}(2)$ part, 
\begin{equation}
\bar{a}_{i}=\frac{1}{2}\sum_{\alpha=1}^{2}a_{i,\alpha},\quad a_{i,\alpha}^{\prime}=a_{i,\alpha}-\bar{a}_{i}.
\end{equation}
From the point of view of an $\mathrm{SU}(2)$ linear quiver gauge theory,
the masses of the anti-fundamental, fundamental and bifundamental
hypermultiplets are given by
\begin{equation}
\bar{\mu}_{\alpha}=a_{0,\alpha}-\bar{a}_{1},\quad\mu_{\alpha}=a_{r+1,\alpha}-\bar{a}_{r},\quad,\mu_{i,i+1}=\bar{a}_{i+1}-\bar{a}_{i},\quad i=1,\cdots,r-1.
\end{equation}
If we identify the Liouville parameter \textbf{$b$ }with the $\Omega$-deformation
parameters $\varepsilon_{1},\varepsilon_{2}$ as
\begin{equation}
b^{2}=\frac{\varepsilon_{1}}{\varepsilon_{2}},
\end{equation}
and relate the conformal dimensions $\Delta_{i}$ with the Coulomb
parameters $\boldsymbol{a}$ in the following way:
\begin{align}
\Delta_{-1} & =  \frac{\varepsilon^{2}-\left(a_{0,1}-a_{0,2}\right)^{2}}{4\varepsilon_{1}\varepsilon_{2}},\quad\Delta_{r+1}=\frac{\varepsilon^{2}-\left(a_{r+1,1}-a_{r+1,2}\right)^{2}}{4\varepsilon_{1}\varepsilon_{2}},\nonumber \\
\Delta_{i} & =  \frac{\left(\bar{a}_{i+1}-\bar{a}_{i}\right)\left(\bar{a}_{i}-\bar{a}_{i+1}+\varepsilon\right)}{4\varepsilon_{1}\varepsilon_{2}},\quad i=0,\cdots,r,
\end{align}
then we have 
\begin{align}
 &  \Bigg\langle V_{\Delta_{-1}}(\infty)\prod_{i=0}^{r}V_{\Delta_{i}}(z_{i})V_{\Delta_{r+1}}(0)\Bigg\rangle\nonumber \\
 & =  f\left(\Delta_{-1},\cdots,\Delta_{r+1}\right)\left|\left(z_{0}^{\Delta_{-1}-\Delta_{0}-\frac{\varepsilon^{2}}{4\varepsilon_{1}\varepsilon_{2}}}\right)\left(\prod_{i=1}^{r-1}z_{i}^{-\Delta_{i}}\right)\left(z_{r}^{\frac{\varepsilon^{2}}{4\varepsilon_{1}\varepsilon_{2}}-\Delta_{r}-\Delta_{r+1}}\right)\right|^{2}\nonumber \\
 &   \times\int\prod_{i=1}^{r}[da_{i}^{\prime}]\left|\frac{\mathcal{Z}\left(\boldsymbol{a};\boldsymbol{z};\varepsilon_{1},\varepsilon_{2}\right)}{\mathcal{Z}^{\mathrm{U}(1)}\left(\boldsymbol{a};\boldsymbol{z};\varepsilon_{1},\varepsilon_{2}\right)}\right|^{2},
\end{align}
where the prefactor $f\left(\Delta_{-1},\cdots,\Delta_{r+1}\right)$
is independent of $\boldsymbol{z}$, and $\mathcal{Z}^{\mathrm{U}(1)}\left(\boldsymbol{a};\boldsymbol{z};\varepsilon_{1},\varepsilon_{2}\right)$
is the $\mathrm{U}(1)$ part of the partition function. 

\subsection{Degenerate partition function}

Up to this point we assumed that the Coulomb moduli $\boldsymbol{a}$
are generic. Then the instanton partition function (\ref{eq:Zinst})
contains an infinite sum over collections of Young diagrams $\boldsymbol{Y}$.
However, we can tune some of the parameters to special values so as
to force some of $Y^{(i,\alpha)}$ to have a constrained shape. For
example, we can adjust
\begin{equation}
a_{0,\alpha}=\begin{cases}
a_{1,1}+(m-1)\varepsilon_{1}+(n-1)\varepsilon_{2}, & \alpha=1,\\
a_{1,\alpha}, & \alpha\neq1,
\end{cases}\label{eq:masstuning}
\end{equation}
where $m,n\in\mathbb{Z}^{+}$. Since the measure of the instanton
partition function contains a factor,
\begin{equation}
\prod_{\square=(u,v)\in Y^{(1,\alpha)}}\left(a_{0,\alpha}-a_{1,\alpha}-\varepsilon_{1}\left(u-1\right)-\varepsilon_{2}\left(v-1\right)\right),
\end{equation}
the contribution to the instanton partition function vanishes unless
the Young diagrams $Y^{(1,\alpha)}=\emptyset$ for $\alpha\neq1$,
and $\square=(m,n)\notin Y^{(1,1)}$. Hence the number of Young diagrams
we need to sum over reduces drastically. In particular, when $m>1$
and $n=1$, the Young diagram $Y^{(1,1)}$ can have at most $m-1$
rows. According to the AGT dictionary, (\ref{eq:masstuning}) corresponds
to a degenerate field with the conformal dimension $\Delta_{(m,n)}$.

Physically, the tuning of the Coulomb parameters (\ref{eq:masstuning}) initiates a partial Higgsing of the four-dimensional $\mathcal{N}=2$ gauge theory, after which only a part of the gauge symmetry is restored on the two-dimensional surfaces $\mathbb{C}_{\varepsilon_1}$ and $\mathbb{C}_{\varepsilon_2}$. Therefore, the partition function of the $\mathcal{N}=2$ gauge theory subject to the constraints yields the correlation function of the surface defects defined by coupling the two-dimensional degrees of freedom on these surfaces to the remaining four-dimensional gauge field. See \cite{Nekrasov:2017rqy, Jeong:2018qpc} for more detail on the Higgsing construction of the surface defects in the $\mathcal{N}=2$ gauge theory.

\section{Superconformal theory with gauge group $\mathrm{U}(N)$ \label{sec:A1}}

In this section, we take a simple example to illustrate the basic
idea of deriving the differential equation on the instanton partition
function at a special point in the parameter space. We consider the
$\mathrm{U}(N)$ gauge theory with $N$ fundamental hypermultiplets and $N$
antifundamental hypermultiplets for general $N\geq2$. At the degenerate
point of parameter space,
\begin{equation}
a_{0,\alpha}=\begin{cases}
a_{1,1}+\varepsilon_{1}, & \alpha=1,\\
a_{1,\alpha}, & \alpha\neq1,
\end{cases}\label{eq:tuning21}
\end{equation}
the instanton partition function is only summed over the Young diagram
$Y^{(1,1)}$ which has only one row,
\begin{equation}
Y^{(1)}=\left(\ytableausetup{mathmode,boxsize=3em, centertableaux}
\begin{ytableau}
(1,1)&(1,2)&\none[\dots]&(1,k_{1})
\end{ytableau},\emptyset,\cdots,\emptyset\right).
\end{equation}
 Therefore, we can label the Young diagram $Y^{(1,1)}$ by the instanton
charge $k_{1}$. 

In this case, we face no obstruction in proving directly that the
instanton partition function is a (generalized) hypergeometric function
from the instanton partition function. The instanton partition function
is
\begin{align}
\mathcal{Z}^{\mathrm{instanton}} & =  \sum_{k_{1}=0}^{\infty}\frac{q_{1}^{k_{1}}}{k_{1}!}\frac{\prod_{\alpha=1}^{N}\left(\frac{a_{0,1}-a_{2,\alpha}+\varepsilon_{2}}{\varepsilon_{2}}\right)^{\overline{k_{1}}}}{\prod_{\alpha=2}^{N}\left(\frac{a_{0,1}-a_{0,\alpha}+\varepsilon_{2}}{\varepsilon_{2}}\right)^{\overline{k_{1}}}}\\
 & =  _{N}F_{N-1}\left(\left(\frac{a_{0,1}-a_{2,\alpha}+\varepsilon_{2}}{\varepsilon_{2}}\right)_{\alpha=1}^{N};\left(\frac{a_{0,1}-a_{0,\alpha}+\varepsilon_{2}}{\varepsilon_{2}}\right)_{\alpha=2}^{N};q_{1}\right),
\end{align}
which is a (generalized) hypergeometric function, and satisfies the
(generalized) hypergeometric differential equation (see the Appendix
\ref{sec:hypergeometric} for details),

\begin{align}
0 & =  \Bigg[q_{1}\prod_{\alpha=1}^{N}\left(q_{1}\frac{\partial}{\partial q_{1}}+\frac{a_{0,1}-a_{2,\alpha}+\varepsilon_{2}}{\varepsilon_{2}}\right)\nonumber \\
 &   -q_{1}\frac{\partial}{\partial q_{1}}\prod_{\alpha=2}^{N}\left(q_{1}\frac{\partial}{\partial q_{1}}+\frac{a_{0,1}-a_{0,\alpha}+\varepsilon_{2}}{\varepsilon_{2}}-1\right)\Bigg]\mathcal{Z}^{\mathrm{instanton}}.\label{eq:DiffA1}
\end{align}

Now we would like to derive the above differential equation using
the NPDS equations. There is only one fundamental qq-character in
this theory,
\begin{equation}
\mathscr{X}_{1}(x)=\mathscr{Y}_{1}(x+\varepsilon)+q_{1}\frac{\mathscr{Y}_{0}(x)\mathscr{Y}_{2}(x+\varepsilon)}{\mathscr{Y}_{1}(x)}.
\end{equation}
At the degenerate point (\ref{eq:tuning21}), the value $\mathscr{Y}_{1}(x)[\boldsymbol{Y}]$
simplifies to
\begin{align}
\mathscr{Y}_{1}(x)[\boldsymbol{Y}] & =  \left[\prod_{\alpha=1}^{N}\left(x-a_{1,\alpha}\right)\right]\left[\prod_{v=1}^{k_{1}}\frac{\left(x-a_{1,1}-\varepsilon_{2}\left(v-1\right)-\varepsilon_{1}\right)\left(x-a_{1,1}-\varepsilon_{2}v\right)}{\left(x-a_{1,1}-\varepsilon_{2}\left(v-1\right)\right)\left(x-a_{1,1}-\varepsilon_{2}v-\varepsilon_{1}\right)}\right]\nonumber \\
 & =  \left[\prod_{\alpha=1}^{N}\left(x-a_{1,\alpha}\right)\right]\left[\frac{\left(x-a_{1,1}-\varepsilon_{1}\right)\left(x-a_{1,1}-\varepsilon_{2}k_{1}\right)}{\left(x-a_{1,1}\right)\left(x-a_{1,1}-\varepsilon_{2}k_{1}-\varepsilon_{1}\right)}\right]\nonumber \\
 & =  \left[\left(x-a_{1,1}-\varepsilon_{1}\right)\prod_{\alpha=2}^{N}\left(x-a_{1,\alpha}\right)\right]\left[\frac{x-a_{1,1}-\varepsilon_{2}k_{1}}{x-a_{1,1}-\varepsilon_{2}k_{1}-\varepsilon_{1}}\right]\nonumber \\
 & =  \mathscr{Y}_{0}(x)\frac{x-a_{0,1}+\varepsilon_{1}-\varepsilon_{2}k_{1}}{x-a_{0,1}-\varepsilon_{2}k_{1}}.\label{eq:Y1simplify}
\end{align}
Accordingly, $\mathscr{X}_{1}(x)[\boldsymbol{Y}]$ becomes
\begin{align}
\mathscr{X}_{1}(x)[\boldsymbol{Y}] & =  \mathscr{Y}_{0}(x+\varepsilon)\left(1+\frac{\varepsilon_{1}}{x+\varepsilon-a_{0,1}-\varepsilon_{2}k_{1}}\right)\nonumber \\
 &   +q_{1}\mathscr{Y}_{2}(x+\varepsilon)\left(1-\frac{\varepsilon_{1}}{x-a_{0,1}+\varepsilon_{1}-\varepsilon_{2}k_{1}}\right).
\end{align}
The $x^{-1}$ coefficient of the large-$x$ expansion of $\mathscr{X}_{1}(x)[\boldsymbol{Y}]$
is given by
\begin{equation}
\mathscr{X}_{1}^{(-1)}[\boldsymbol{Y}]=\varepsilon_{1}\mathscr{Y}_{0}(a_{0,1}+\varepsilon_{2}k_{1})-q_{1}\varepsilon_{1}\mathscr{Y}_{2}(a_{0,1}+\varepsilon_{2}k_{1}+\varepsilon_{2}).
\end{equation}
Using the relation
\begin{align}
\langle k_{1}^{p}\rangle & =  \mathcal{Z}^{\mathrm{instanton}}\left(\boldsymbol{a};\vec{Y};\varepsilon_{1},\varepsilon_{2}\right)^{-1}\sum_{\vec{Y}=\left\{ Y^{(\alpha)}\right\} }q_{1}^{k_{1}}\mathcal{Z}^{\mathrm{instanton}}\left(\boldsymbol{a};\vec{Y};\varepsilon_{1},\varepsilon_{2}\right)k_{1}^{p}\nonumber \\
 & =  \mathcal{Z}^{\mathrm{instanton}}\left(\boldsymbol{a};\vec{Y};\varepsilon_{1},\varepsilon_{2}\right)^{-1}\left(q_{1}\frac{\partial}{\partial q_{1}}\right)^{p}\mathcal{Z}^{\mathrm{instanton}}\left(\boldsymbol{a};q_{1};\varepsilon_{1},\varepsilon_{2}\right),
\end{align}
the equation $\langle\mathscr{X}_{1}^{(-1)}\rangle=0$ becomes
\begin{equation}
0=\left[\prod_{\alpha=1}^{N}\left(a_{0,1}+\varepsilon_{2}q_{1}\frac{\partial}{\partial q_{1}}-a_{0,\alpha}\right)-q_{1}\prod_{\alpha=1}^{N}\left(a_{0,1}+\varepsilon_{2}q_{1}\frac{\partial}{\partial q_{1}}+\varepsilon_{2}-a_{2,\alpha}\right)\right]\mathcal{Z}^{\mathrm{instanton}},
\end{equation}
which coincides with the differential equation (\ref{eq:DiffA1}).

\section{Superconformal linear quiver gauge theories \label{sec:quiver}}

In this section, we would like to derive the differential equation
on the instanton partition function of the superconformal linear quiver
gauge theory using the NPDS equations.

\subsection{Large-$x$ expansion of fundamental \textmd{\normalsize{}$\mathscr{Y}$-}observables}

The first step is to compute the large-$x$ expansion of the $\mathscr{Y}$-observables,
\begin{align}
\mathscr{Y}_{i}(x)[\boldsymbol{Y}] & =  x^{N}\exp\left[\sum_{\alpha=1}^{N}\log\left(1-\frac{a_{i,\alpha}}{x}\right)+\sum_{\alpha=1}^{N}\sum_{\square\in Y^{(i,\alpha)}}\frac{\left(1-\frac{\widehat{c_{\square}}+\varepsilon_{1}}{x}\right)\left(1-\frac{\widehat{c_{\square}}+\varepsilon_{2}}{x}\right)}{\left(1-\frac{\widehat{c_{\square}}}{x}\right)\left(1-\frac{\widehat{c_{\square}}+\varepsilon}{x}\right)}\right]\nonumber \\
 & =  x^{N}\exp\left(-\sum_{n=1}^{\infty}\frac{C_{i,n}[\boldsymbol{Y}]}{nx^{n}}\right),
\end{align}
where 
\begin{align}
C_{i,n}[\boldsymbol{Y}] & =  \mathrm{Tr}\Phi_{i}^{n}(0)[\boldsymbol{Y}]\nonumber \\
 & = \sum_{\alpha=1}^{N}\left\{ a_{i,\alpha}^{n}+\sum_{\square\in Y^{(i,\alpha)}}\left[\left(\widehat{c_{\square}}+\varepsilon_{1}\right)^{n}+\left(\widehat{c_{\square}}+\varepsilon_{2}\right)^{n}-\widehat{c_{\square}}^{n}-\left(\widehat{c_{\square}}+\varepsilon\right)^{n}\right]\right\} .
\end{align}
In particular, we have
\begin{align}
C_{i,1}[\boldsymbol{Y}] & =  \sum_{\alpha=1}^{N}a_{i,\alpha},\nonumber \\
C_{i,2}[\boldsymbol{Y}] & =  \left(\sum_{\alpha=1}^{N}a_{i,\alpha}^{2}\right)-2\varepsilon_{1}\varepsilon_{2}k_{i},
\end{align}
We also have the similar expression for $\mathscr{Y}_{i}(x+\varepsilon)[\boldsymbol{Y}]$,
\begin{align}
\mathscr{Y}_{i}(x+\varepsilon)[\boldsymbol{Y}] & =  \prod_{\alpha=1}^{N}\left(x-\left(a_{i,\alpha}-\varepsilon\right)\right)\prod_{\square\in Y^{(i,\alpha)}}\frac{\left(x-\widehat{c_{\square}}+\varepsilon_{1}\right)\left(x-\widehat{c_{\square}}+\varepsilon_{2}\right)}{\left(x-\widehat{c_{\square}}\right)\left(x-\widehat{c_{\square}}+\varepsilon\right)}\nonumber \\
 & =  x^{N}\exp\left(-\sum_{n=1}^{\infty}\frac{C_{i,n}^{\prime}[\boldsymbol{Y}]}{nx^{n}}\right),
\end{align}
where 
\begin{equation}
C_{i,n}^{\prime}[\boldsymbol{Y}]=\sum_{\alpha=1}^{N}\left\{ \left(a_{i,\alpha}-\varepsilon\right)^{n}+\sum_{\square\in Y^{(i,\alpha)}}\left[\left(\widehat{c_{\square}}-\varepsilon_{1}\right)^{n}+\left(\widehat{c_{\square}}-\varepsilon_{2}\right)^{n}-\widehat{c_{\square}}^{n}-\left(\widehat{c_{\square}}-\varepsilon\right)^{n}\right]\right\} .
\end{equation}
Therefore, we obtain the large-$x$ expansion of $\Xi_{i}(x)[\boldsymbol{Y}]$,
\begin{equation}
\Xi_{i}(x)[\boldsymbol{Y}]=\frac{\mathscr{Y}_{i+1}(x+\varepsilon)[\boldsymbol{Y}]}{\mathscr{Y}_{i}(x)[\boldsymbol{Y}]}=1+\sum_{n=1}^{\infty}\frac{\zeta_{i,n}[\boldsymbol{Y}]}{x^{n}}.\label{eq:XiY}
\end{equation}
The first two terms of $\zeta_{i,n}$ are given explicitly as 
\begin{align}
\zeta_{i,1} & =  \mathcal{A}_{i}^{(1)},\nonumber \\
\zeta_{i,2} & = \mathcal{A}_{i}^{(2)}-\varepsilon_{1}\varepsilon_{2}\left(k_{i}-k_{i+1}\right),
\end{align}
where
\begin{align}
\mathcal{A}_{i}^{(1)} & =  \sum_{\alpha=1}^{N}\left(a_{i,\alpha}-a_{i+1,\alpha}+\varepsilon\right),\nonumber \\
\mathcal{A}_{i}^{(2)} & =  \frac{1}{2}\sum_{\alpha=1}^{N}\left[a_{i,\alpha}^{2}-\left(a_{i+1,\alpha}-\varepsilon\right)^{2}\right]+\frac{1}{2}\left(\mathcal{A}_{i}^{(1)}\right)^{2}
\end{align}

\subsection{Generating function of the fundamental qq-characters}

After expanding the $\mathscr{Y}$-observables, we would like to calculate
the large-$x$ expansion of the qq-characters. In order to deal with
all of the fundamental qq-characters at the same time we introduce
the generating function,
\begin{align}
\mathscr{G}_{r}(x;t) & =  \mathscr{Y}_{0}(x)^{-1}\Delta_{r}^{-1}\sum_{\ell=0}^{r+1}z_{0}z_{1}\cdots z_{\ell-1}t^{\ell}\mathscr{X}_{\ell}\left(x-\varepsilon(1-\ell)\right)\nonumber \\
 & =  \Delta_{r}^{-1}\sum_{I\subset[0,r]}\left[\left(\prod_{i\in I}tz_{i}\right)\prod_{i\in I}\Xi_{i}\left(x+\varepsilon h_{I}(i)\right)\right],\label{eq:Gr}
\end{align}
where 
\begin{equation}
\Delta_{r}=\sum_{I\subset[0,r]}\left(\prod_{i\in I}tz_{i}\right)=\prod_{i=0}^{r}\left(1+tz_{i}\right).
\end{equation}
In the following, we would like to sum over $I\subset[0,r]$ to obtain
the large-$x$ expansion of $\mathscr{G}_{r}(x;t)$,
\begin{equation}
\mathscr{G}_{r}(x;t)=\sum_{n=0}^{\infty}\frac{\mathscr{G}_{r}^{(-n)}(t)}{x^{n}}.
\end{equation}

Let us define
\begin{equation}
u_{i}=\frac{tz_{i}}{1+tz_{i}}.
\end{equation}
When $r=0$, $\mathscr{G}_{0}(t)$ is given by a sum over $I=\emptyset$
and $I=\left\{ 0\right\} $, 
\begin{equation}
\mathscr{G}_{0}(x;t)=\frac{1}{1+tz_{0}}+\frac{tz_{0}}{1+tz_{0}}\Xi_{0}(x)=1+\sum_{n=1}^{\infty}\frac{u_{0}\zeta_{0,n}}{x^{n}}.
\end{equation}
Hence,
\begin{equation}
\mathscr{G}_{0}^{(0)}(t)=1,\quad\mathscr{G}_{0}^{(-n)}(t)=u_{0}\zeta_{0,n},\quad n\in\mathbb{Z}^{+}.
\end{equation}

For general $r\geq1$, we can compute the value of the generating
function (\ref{eq:Gr}) using the recurrence relation between $\mathscr{G}_{r}(x;t)$
and $\mathscr{G}_{r-1}\left(x;t\right)$. We divide the sum over $I\subset\left[0,r\right]$
into two classes: $r\notin I$ and $r\in I$, 
\begin{align}
\mathscr{G}_{r}(x;t) & =  \frac{1}{1+tz_{r}}\Delta_{r-1}^{-1}\sum_{I^{\prime}\subset[0,r-1]}\left[\left(\prod_{i\in I^{\prime}}tz_{i}\right)\prod_{i\in I^{\prime}}\Xi_{i}\left(x+\varepsilon h_{I^{\prime}}(i)\right)\right]\nonumber \\
 &   +\frac{tz_{r}}{1+tz_{r}}\Delta_{r-1}^{-1}\sum_{I^{\prime}\subset[0,r-1]}\left[\left(\prod_{i\in I^{\prime}}tz_{i}\right)\Xi_{r}\left(x+\varepsilon|I^{\prime}|\right)\prod_{i\in I^{\prime}}\Xi_{i}\left(x+\varepsilon h_{I^{\prime}}(i)\right)\right]\nonumber \\
 & =  \mathscr{G}_{r-1}(x;t)+u_{r}\Delta_{r-1}^{-1}\left(\Xi_{r}\left(x+\varepsilon t\frac{\partial}{\partial t}\right)-1\right)\left(\Delta_{r-1}\mathscr{G}_{r-1}(x;t)\right)\nonumber \\
 & =  \mathscr{G}_{r-1}(x;t)+u_{r}\Delta_{r-1}^{-1}\sum_{n=1}^{\infty}\frac{\zeta_{r,n}}{\left(x+\varepsilon t\frac{\partial}{\partial t}\right)^{n}}\left(\Delta_{r-1}\mathscr{G}_{r-1}(x;t)\right)\nonumber \\
 & =  \mathscr{G}_{0}(x;t)+\sum_{j=1}^{r}u_{j}\Delta_{j-1}^{-1}\sum_{n=1}^{\infty}\frac{\zeta_{j,n}}{\left(x+\varepsilon t\frac{\partial}{\partial t}\right)^{n}}\left(\Delta_{j-1}\mathscr{G}_{j-1}(x;t)\right).
\end{align}
Hence, we obtain the recursive relations, 
\begin{align}
\mathscr{G}_{r}^{(0)}(t) & = \mathscr{G}_{0}^{(0)}(t)=1,\\
\mathscr{G}_{r}^{(-1)}(t) & =  \sum_{j=0}^{r}u_{j}\zeta_{j,1},\\
\mathscr{G}_{r}^{(-2)}(t) & =  \sum_{j=0}^{r}u_{j}\zeta_{j,2}+\sum_{j=1}^{r}u_{j}\zeta_{j,1}\mathscr{G}_{j-1}^{(-1)}(t)\nonumber \\
 &   -\varepsilon\sum_{j=1}^{r}u_{j}\zeta_{j,1}\Delta_{j-1}^{-1}t\frac{\partial}{\partial t}\Delta_{j-1},\\
\mathscr{G}_{r}^{(-3)}(t) & =  \sum_{j=0}^{r}u_{j}\zeta_{j,3}+\sum_{j=1}^{r}u_{j}\left[\zeta_{j,1}\mathscr{G}_{j-1}^{(-2)}(t)+\zeta_{j,2}\mathscr{G}_{j-1}^{(-1)}(t)\right]\nonumber \\
 &   -\varepsilon\sum_{j=1}^{r}u_{j}\Delta_{j-1}^{-1}\left[\zeta_{j,1}t\frac{\partial}{\partial t}\left(\Delta_{j-1}\mathscr{G}_{j-1}^{(-1)}(t)\right)+2\zeta_{j,2}t\frac{\partial}{\partial t}\Delta_{j-1}\right]\nonumber \\
 &   +\varepsilon^{2}\sum_{j=1}^{r}u_{j}\zeta_{j,1}\Delta_{j-1}^{-1}\left(t\frac{\partial}{\partial t}\right)^{2}\Delta_{j-1},\\
\mathscr{G}_{r}^{(-4)}(t) & =  \sum_{j=0}^{r}u_{j}\zeta_{j,4}+\sum_{j=1}^{r}u_{j}\left[\zeta_{j,1}\mathscr{G}_{j-1}^{(-3)}(t)+\zeta_{j,2}\mathscr{G}_{j-1}^{(-2)}(t)+\zeta_{j,3}\mathscr{G}_{j-1}^{(-1)}(t)\right]\nonumber \\
 &   -\varepsilon\sum_{j=1}^{r}u_{j}\Delta_{j-1}^{-1}\left[\zeta_{j,1}t\frac{\partial}{\partial t}\left(\Delta_{j-1}\mathscr{G}_{j-1}^{(-2)}(t)\right)+2\zeta_{j,2}t\frac{\partial}{\partial t}\left(\Delta_{j-1}\mathscr{G}_{j-1}^{(-1)}(t)\right)+3\zeta_{j,3}t\frac{\partial}{\partial t}\Delta_{j-1}\right]\nonumber \\
 &   +\varepsilon^{2}\sum_{j=1}^{r}u_{j}\Delta_{j-1}^{-1}\left[\zeta_{j,1}\left(t\frac{\partial}{\partial t}\right)^{2}\left(\Delta_{j-1}\mathscr{G}_{j-1}^{(-1)}(t)\right)+3\zeta_{j,2}\left(t\frac{\partial}{\partial t}\right)^{2}\Delta_{j-1}\right]\nonumber \\
 &   -\varepsilon^{3}\sum_{j=1}^{r}u_{j}\zeta_{j,1}\Delta_{j-1}^{-1}\left(t\frac{\partial}{\partial t}\right)^{3}\Delta_{j-1}.
\end{align}
We further introduce the notation
\begin{equation}
U_{r}\left[s_{1},s_{2},\cdots,s_{\ell}\right]\equiv\sum_{0\leq i_{1}<\cdots<i_{\ell}\leq r}\prod_{n=1}^{\ell}\left(u_{i_{n}}\zeta_{i_{n},s_{n}}\right),
\end{equation}
where $\left[s_{1},\cdots,s_{\ell}\right]$ is a sequence of non-negative
integers, and we adopt the convention that $\zeta_{i,0}=1$. We have
the following useful relations from the definition,
\begin{align}
t\frac{\partial}{\partial t}\Delta_{r} & =  \Delta_{r}U_{r}[0],\\
t\frac{\partial}{\partial t}\left(\Delta_{r}U_{r}[s_{1},\cdots,s_{\ell}]\right) & =  \Delta_{r}\left(\ell U_{r}[s_{1},\cdots,s_{\ell}]+U_{r}^{\oplus}[s_{1},\cdots,s_{\ell}]\right),\\
\sum_{j=1}^{r}u_{j}\zeta_{j,m}U_{j-1}[s_{1},\cdots,s_{\ell}] & =  U_{r}\left[s_{1},s_{2},\cdots,s_{\ell},m\right],
\end{align}
where
\begin{equation}
U_{r}^{\oplus}[s_{1},\cdots,s_{\ell}]\equiv U_{r}[0,s_{1},\cdots,s_{\ell}]+U_{r}[s_{1},0,\cdots,s_{\ell}]+\cdots+U_{r}[s_{1},\cdots,s_{\ell},0].
\end{equation}
We also have 
\begin{align}
\Delta_{r}^{-1}t\frac{\partial}{\partial t}\Delta_{r} & =  U_{r}[0],\\
\Delta_{r}^{-1}\left(t\frac{\partial}{\partial t}\right)^{2}\Delta_{r} & =  U_{r}[0]+2U_{r}[0,0],\\
\Delta_{r}^{-1}\left(t\frac{\partial}{\partial t}\right)^{3}\Delta_{r} & = U_{r}[0]+6U_{r}[0,0]+6U_{r}[0,0,0],
\end{align}
After solving the recurrence relations, the first few terms of $\mathscr{G}_{r}^{(-n)}(t)$
can be written as
\begin{align}
\mathscr{G}_{r}^{(0)}(t) & =  1,\\
\mathscr{G}_{r}^{(-1)}(t) & =  U_{r}[1],\\
\mathscr{G}_{r}^{(-2)}(t) & =  U_{r}[2]+U_{r}[1,1]-\varepsilon U_{r}[0,1],\\
\mathscr{G}_{r}^{(-3)}(t) & =  U_{r}[3]+U_{r}[2,1]+U_{r}[1,2]-\varepsilon\left(U_{r}[1,1]+2U_{r}[0,2]\right)+\varepsilon^{2}U_{r}[0,1]\nonumber \\
 &   +U_{r}[1,1,1]-\varepsilon\left(2U_{r}[0,1,1]+U_{r}[1,0,1]\right)+2\varepsilon^{2}U_{r}[0,0,1],\\
\mathscr{G}_{r}^{(-4)}(t) & =  U_{r}[4]+U_{r}[1,3]+U_{r}[2,2]+U_{r}[3,1]\nonumber \\
 &   -\varepsilon\left(U_{r}[2,1]+2U_{r}[1,2]+3U_{r}[0,3]\right)+\varepsilon^{2}\left(U_{r}[1,1]+U_{r}[0,2]\right)-\varepsilon^{3}U_{r}[0,1]\nonumber \\
 &   +U_{r}[2,1,1]+U_{r}[1,2,1]+U_{r}[1,1,2]\nonumber \\
 &  -\varepsilon\left(3U_{r}[1,1,1]+3U_{r}[0,2,1]+3U_{r}[0,1,2]+2U_{r}[1,0,2]+U_{r}[2,0,1]\right)\nonumber \\
 &   +\varepsilon^{2}\left(6U_{r}[0,1,1]+3U_{r}[1,0,1]+6U_{r}[0,0,2]\right)-6\varepsilon^{3}U_{r}[0,0,1]\nonumber \\
 &   +U_{r}[1,1,1,1]-\varepsilon\left(3U_{r}[0,1,1,1]+2U_{r}[1,0,1,1]+U_{r}[1,1,0,1]\right)\nonumber \\
 &   +\varepsilon^{2}\left(6U_{r}[0,0,1,1]+3U_{r}[0,1,0,1]+2U_{r}[1,0,0,1]\right)-6\varepsilon^{3}U_{r}[0,0,0,1].
\end{align}

In this paper, we are interested in the special case $N=2$, with
$\mathscr{Y}_{0}(x)=x^{2}-\left(a_{0,1}+a_{0,2}\right)x+a_{0,1}a_{0,2}$.
We can deduce from the NPDS equations (\ref{eq:npDS}) that $\langle\mathscr{Y}_{0}(x)\mathscr{G}_{r}(x;t)\rangle$
is a polynomial in $x$ for arbitrary $t$. In particular, we have
\begin{align}
0 & =  \langle\mathscr{G}_{r}^{(-3)}(t)\rangle-\left(a_{0,1}+a_{0,2}\right)\langle\mathscr{G}_{r}^{(-2)}(t)\rangle+a_{0,1}a_{0,2}\langle\mathscr{G}_{r}^{(-1)}(t)\rangle\nonumber \\
 & =  \langle U_{r}[3]\rangle-\left(a_{0,1}+a_{0,2}\right)\langle U_{r}[2]\rangle+a_{0,1}a_{0,2}\langle U_{r}[1]\rangle\nonumber \\
 &   +\langle U_{r}[2,1]\rangle+\langle U_{r}[1,2]\rangle-2\varepsilon\langle U_{r}[0,2]\rangle-\left(a_{0,1}+a_{0,2}+\varepsilon\right)\langle U_{r}[1,1]\rangle+\varepsilon\left(a_{0,1}+a_{0,2}+\varepsilon\right)\langle U_{r}[0,1]\rangle\nonumber \\
 &   +\langle U_{r}[1,1,1]\rangle-\varepsilon\langle U_{r}[1,0,1]\rangle-2\varepsilon\langle U_{r}[0,1,1]\rangle+2\varepsilon^{2}\langle U_{r}[0,0,1]\rangle,\label{eq:NPDS-1}
\end{align}
and
\begin{align}
0 & =  \langle\mathscr{G}_{r}^{(-4)}(t)\rangle-\left(a_{0,1}+a_{0,2}\right)\langle\mathscr{G}_{r}^{(-3)}(t)\rangle+a_{0,1}a_{0,2}\langle\mathscr{G}_{r}^{(-2)}(t)\rangle\nonumber \\
 & =  \langle U_{r}[4]\rangle-\left(a_{0,1}+a_{0,2}\right)\langle U_{r}[3]\rangle+a_{0,1}a_{0,2}\langle U_{r}[2]\rangle\nonumber \\
 &   +\langle U_{r}[1,3]\rangle+\langle U_{r}[3,1]\rangle-3\varepsilon\langle U_{r}[0,3]\rangle+\langle U_{r}[2,2]\rangle\nonumber \\
 &   -\left(a_{0,1}+a_{0,2}+\varepsilon\right)\langle U_{r}[2,1]\rangle-\left(a_{0,1}+a_{0,2}+2\varepsilon\right)\langle U_{r}[1,2]\rangle+\varepsilon\left(\varepsilon+2a_{0,1}+2a_{0,2}\right)\langle U_{r}[0,2]\rangle\nonumber \\
 &  +\left(a_{0,1}+\varepsilon\right)\left(a_{0,2}+\varepsilon\right)\langle U_{r}[1,1]\rangle-\varepsilon\left(a_{0,1}+\varepsilon\right)\left(a_{0,2}+\varepsilon\right)\langle U_{r}[0,1]\rangle\nonumber \\
 &   +\langle U_{r}[2,1,1]\rangle+\langle U_{r}[1,2,1]\rangle+\langle U_{r}[1,1,2]\rangle\nonumber \\
 &   -\varepsilon\langle U_{r}[2,0,1]\rangle-2\varepsilon\langle U_{r}[1,0,2]\rangle-3\varepsilon\langle U_{r}[0,2,1]\rangle-3\varepsilon\langle U_{r}[0,1,2]\rangle+6\varepsilon^{2}\langle U_{r}[0,0,2]\rangle\nonumber \\
 &   -\left(a_{0,1}+a_{0,2}+3\varepsilon\right)\langle U_{r}[1,1,1]\rangle+\varepsilon\left(a_{0,1}+a_{0,2}+3\varepsilon\right)\langle U_{r}[1,0,1]\rangle\nonumber \\
 &   +2\varepsilon\left(a_{0,1}+a_{0,2}+3\varepsilon\right)\langle U_{r}[0,1,1]\rangle-2\varepsilon^{2}\left(a_{0,1}+a_{0,2}+3\varepsilon\right)\langle U_{r}[0,0,1]\rangle\nonumber \\
 &   +\langle U_{r}[1,1,1,1]\rangle-\varepsilon\left(3\langle U_{r}[0,1,1,1]\rangle+2\langle U_{r}[1,0,1,1]\rangle+\langle U_{r}[1,1,0,1]\rangle\right)\nonumber \\
 &   +\varepsilon^{2}\left(6\langle U_{r}[0,0,1,1]\rangle+3\langle U_{r}[0,1,0,1]\rangle+2\langle U_{r}[1,0,0,1]\rangle\right)-6\varepsilon^{3}\langle U_{r}[0,0,0,1]\rangle.\label{eq:NPDS-2}
\end{align}
By taking the residue of (\ref{eq:NPDS-1}) at $t=-z_{i}^{-1}$, we
have
\begin{align}
0 & =  \langle\zeta_{i,3}\rangle+\left[-a_{0,1}-a_{0,2}+\sum_{j=0}^{i-1}\frac{z_{j}}{z_{j}-z_{i}}\left(\mathcal{A}_{j}^{(1)}-2\varepsilon\right)+\sum_{j=i+1}^{r}\frac{z_{j}}{z_{j}-z_{i}}\mathcal{A}_{j}^{(1)}\right]\langle\zeta_{i,2}\rangle+a_{0,1}a_{0,2}\mathcal{A}_{i}^{(1)}\nonumber \\
 &   +\sum_{j=0}^{i-1}\frac{z_{j}}{z_{j}-z_{i}}\left[\mathcal{A}_{i}^{(1)}\langle\zeta_{j,2}\rangle-\left(a_{0,1}+a_{0,2}+\varepsilon\right)\left(\mathcal{A}_{j}^{(1)}-\varepsilon\right)\mathcal{A}_{i}^{(1)}\right]\nonumber \\
 &   +\sum_{j=i+1}^{r}\frac{z_{j}}{z_{j}-z_{i}}\left[\left(\mathcal{A}_{i}^{(1)}-2\varepsilon\right)\langle\zeta_{j,2}\rangle-\left(a_{0,1}+a_{0,2}+\varepsilon\right)\left(\mathcal{A}_{i}^{(1)}-\varepsilon\right)\mathcal{A}_{j}^{(1)}\right]\nonumber \\
 &   +\sum_{0\leq i_{1}<i_{2}<i}\frac{z_{i_{1}}z_{i_{2}}}{\left(z_{i_{1}}-z_{i}\right)\left(z_{i_{2}}-z_{i}\right)}\left(\mathcal{A}_{i_{1}}^{(1)}-2\varepsilon\right)\left(\mathcal{A}_{i_{2}}^{(1)}-\varepsilon\right)\mathcal{A}_{i}^{(1)}\nonumber \\
 &   +\sum_{i_{1}=0}^{i-1}\sum_{i_{2}=i+1}^{r}\frac{z_{i_{1}}z_{i_{2}}}{\left(z_{i_{1}}-z_{i}\right)\left(z_{i_{2}}-z_{i}\right)}\left(\mathcal{A}_{i_{1}}^{(1)}-2\varepsilon\right)\left(\mathcal{A}_{i}^{(1)}-\varepsilon\right)\mathcal{A}_{i_{2}}^{(1)}\nonumber \\
 &   +\sum_{i<i_{1}<i_{2}\leq r}\frac{z_{i_{1}}z_{i_{2}}}{\left(z_{i_{1}}-z_{i}\right)\left(z_{i_{2}}-z_{i}\right)}\left(\mathcal{A}_{i}^{(1)}-2\varepsilon\right)\left(\mathcal{A}_{i_{1}}^{(1)}-\varepsilon\right)\mathcal{A}_{i_{2}}^{(1)}.\label{eq:NPDS-1j}
\end{align}
In particular, when $j=0$, we have
\begin{align}
0 & =  \langle\zeta_{0,3}\rangle-\left(a_{0,1}+a_{0,2}+\sum_{i=1}^{r}\frac{z_{i}}{z_{0}-z_{i}}\mathcal{A}_{i}^{(1)}\right)\langle\zeta_{0,2}\rangle+a_{0,1}a_{0,2}\mathcal{A}_{0}^{(1)}\nonumber \\
 &  -\sum_{i=1}^{r}\frac{z_{i}}{z_{0}-z_{i}}\left[\left(\mathcal{A}_{0}^{(1)}-2\varepsilon\right)\langle\zeta_{i,2}\rangle-\left(a_{0,1}+a_{0,2}+\varepsilon\right)\left(\mathcal{A}_{0}^{(1)}-\varepsilon\right)\mathcal{A}_{i}^{(1)}\right]\nonumber \\
 &   +\sum_{1\leq i_{1}<i_{2}\leq r}\frac{z_{i_{1}}z_{i_{2}}}{\left(z_{0}-z_{i_{1}}\right)\left(z_{0}-z_{i_{2}}\right)}\left(\mathcal{A}_{0}^{(1)}-2\varepsilon\right)\left(\mathcal{A}_{i_{1}}^{(1)}-\varepsilon\right)\mathcal{A}_{i_{2}}^{(1)}.\label{eq:NPDS-10}
\end{align}
We also need the equation obtained by taking residue of (\ref{eq:NPDS-2})
at $t=-z_{0}^{-1}$,
\begin{align}
0 & =  \langle\zeta_{0,4}\rangle-\left(a_{0,1}+a_{0,2}+\sum_{i=1}^{r}\frac{z_{i}}{z_{0}-z_{i}}\mathcal{A}_{i}^{(1)}\right)\langle\zeta_{0,3}\rangle+a_{0,1}a_{0,2}\langle\zeta_{0,2}\rangle\nonumber \\
 &   -\sum_{i=1}^{r}\frac{z_{i}}{z_{0}-z_{i}}\Bigg[\left(\mathcal{A}_{0}^{(1)}-3\varepsilon\right)\langle\zeta_{i,3}\rangle+\langle\zeta_{0,2}\zeta_{i,2}\rangle-\left(a_{0,1}+a_{0,2}+\varepsilon\right)\mathcal{A}_{i}^{(1)}\langle\zeta_{0,2}\rangle\nonumber \\
 &   +\left(\varepsilon\left(\varepsilon+2a_{0,1}+2a_{0,2}\right)-\left(a_{0,1}+a_{0,2}+2\varepsilon\right)\mathcal{A}_{0}^{(1)}\right)\langle\zeta_{i,2}\rangle+\left(a_{0,1}+\varepsilon\right)\left(a_{0,2}+\varepsilon\right)\left(\mathcal{A}_{0}^{(1)}-\varepsilon\right)\mathcal{A}_{i}^{(1)}\Bigg]\nonumber \\
 &   +\sum_{1\leq i_{1}<i_{2}\leq r}\frac{z_{i_{1}}z_{i_{2}}}{\left(z_{0}-z_{i_{1}}\right)\left(z_{0}-z_{i_{2}}\right)}\Bigg[\left(\mathcal{A}_{i_{1}}^{(1)}-\varepsilon\right)\mathcal{A}_{i_{2}}^{(1)}\langle\zeta_{0,2}\rangle+\left(\mathcal{A}_{0}^{(1)}-3\varepsilon\right)\mathcal{A}_{i_{2}}^{(1)}\langle\zeta_{i_{1},2}\rangle\nonumber \\
 &   +\left(\mathcal{A}_{0}^{(1)}-3\varepsilon\right)\left(\mathcal{A}_{i_{1}}^{(1)}-2\varepsilon\right)\langle\zeta_{i_{2},2}\rangle-\left(a_{0,1}+a_{0,2}+3\varepsilon\right)\left(\mathcal{A}_{0}^{(1)}-2\varepsilon\right)\left(\mathcal{A}_{i_{1}}^{(1)}-\varepsilon\right)\mathcal{A}_{i_{2}}^{(1)}\Bigg]\nonumber \\
 &   -\sum_{1\leq i_{1}<i_{2}<i_{3}\leq r}\frac{z_{i_{1}}z_{i_{2}}z_{i_{3}}}{\left(z_{0}-z_{i_{1}}\right)\left(z_{0}-z_{i_{2}}\right)\left(z_{0}-z_{i_{3}}\right)}\left(\mathcal{A}_{0}^{(1)}-3\varepsilon\right)\left(\mathcal{A}_{i_{1}}^{(1)}-2\varepsilon\right)\left(\mathcal{A}_{i_{2}}^{(1)}-\varepsilon\right)\mathcal{A}_{i_{3}}^{(1)}.\label{eq:NPDS-20}
\end{align}

\subsection{Derivation of the differential equations}

Now we are ready to derive the differential equations satisfied by
the instanton partition function using the NPDS equations. The key
point is that $\zeta_{0,n}$ take special values at a degenerate point
in the parameter space. 

\subsubsection{Second-order differential equation}

In order to derive a second-order differential equation, we should
tune the parameters in the following way,
\begin{equation}
a_{0,1}=a_{1,1}+\varepsilon_{1},\quad a_{0,2}=a_{1,2}.
\end{equation}
The configuration of the gauge fields are constrained so that the
Young diagram $Y^{(1,1)}$ has only one row and $Y^{(1,2)}=\emptyset$,
\begin{equation}
Y^{(1)}=\left(\ytableausetup{mathmode,boxsize=3em, centertableaux}
\begin{ytableau}(1,1)&(1,2)&\none[\dots]&(1,k_{1})
\end{ytableau},\emptyset\right).
\end{equation}
 Hence, the Young diagram $Y^{(1,1)}$ is completely determined by
the instanton charge $k_{1}$, and
\begin{align}
\Xi_{0}(x)[k_{1}] & =  \frac{\mathscr{Y}_{1}(x+\varepsilon)[k_{1}]}{\mathscr{Y}_{0}(x)}\nonumber \\
 & =  \frac{\left(x+\varepsilon-a_{0,1}\right)\left(x+\varepsilon-a_{0,2}\right)}{\left(x-a_{0,1}\right)\left(x-a_{0,2}\right)}\frac{x-a_{0,1}+2\varepsilon_{1}-\varepsilon_{2}\left(k_{1}-1\right)}{x-a_{0,1}+\varepsilon_{1}-\varepsilon_{2}\left(k_{1}-1\right)}\nonumber \\
 & =  1+\sum_{n=1}^{\infty}\frac{\zeta_{0,n}[k_{1}]}{x^{n}},
\end{align}
which gives
\begin{align}
\zeta_{0,1}[k_{1}] & =  3\varepsilon_{1}+2\varepsilon_{2},\nonumber \\
\zeta_{0,2}[k_{1}] & =  \left(2\varepsilon_{1}+\varepsilon_{2}\right)a_{0,1}+\varepsilon a_{0,2}+\varepsilon\left(2\varepsilon_{1}+\varepsilon_{2}\right)+\varepsilon_{1}\varepsilon_{2}k_{1},\nonumber \\
\zeta_{0,3}[k_{1}] & =  \left(2\varepsilon_{1}+\varepsilon_{2}\right)a_{0,1}^{2}+\varepsilon\left(2\varepsilon_{1}+\varepsilon_{2}\right)a_{0,1}+\varepsilon a_{0,2}^{2}+\varepsilon\left(2\varepsilon_{1}+\varepsilon_{2}\right)a_{0,2}\nonumber \\
 &   +2\varepsilon_{1}\varepsilon_{2}a_{0,1}k_{1}+\varepsilon_{1}\varepsilon_{2}^{2}k_{1}^{2}.
\end{align}
Hence, from (\ref{eq:NPDS-10}), we have
\begin{align}
0 & =  \varepsilon_{1}\varepsilon_{2}\left(a_{0,1}-a_{0,2}\right)\langle k_{1}\rangle+\varepsilon_{1}\varepsilon_{2}^{2}\langle k_{1}^{2}\rangle\nonumber \\
 &   -\sum_{i=1}^{r}\frac{z_{i}}{z_{0}-z_{i}}\varepsilon_{1}\left[-a_{0,2}\mathcal{A}_{i}^{(1)}+\varepsilon_{2}\mathcal{A}_{i}^{(1)}\langle k_{1}\rangle+\mathcal{A}_{i}^{(2)}-\varepsilon_{1}\varepsilon_{2}\langle k_{i}-k_{i+1}\rangle\right]\nonumber \\
 &   +\sum_{1\leq i_{1}<i_{2}\leq r}\frac{z_{i_{1}}z_{i_{2}}}{\left(z_{0}-z_{i_{1}}\right)\left(z_{0}-z_{i_{2}}\right)}\varepsilon_{1}\left(\mathcal{A}_{i_{1}}^{(1)}-\varepsilon\right)\mathcal{A}_{i_{2}}^{(1)}.
\end{align}
Using
\begin{equation}
\mathcal{Z}^{\mathrm{instanton}}\langle k_{1}\rangle=-\nabla_{0}\mathcal{Z}^{\mathrm{instanton}},\quad\mathcal{Z}^{\mathrm{instanton}}\langle k_{i}-k_{i+1}\rangle=\nabla_{i}\mathcal{Z}^{\mathrm{instanton}},
\end{equation}
we obtain a differential equation on the instanton partition function,
\begin{align}
0 & =  \Bigg\{\varepsilon_{2}^{2}\nabla_{0}^{2}-\varepsilon_{2}\left(a_{0,1}-a_{0,2}-\sum_{i=1}^{r}\frac{z_{i}}{z_{0}-z_{i}}\mathcal{A}_{i}^{(1)}\right)\nabla_{0}\nonumber \\
 &   +\sum_{i=1}^{r}\frac{z_{i}}{z_{0}-z_{i}}\left[a_{0,2}\mathcal{A}_{i}^{(1)}-\mathcal{A}_{i}^{(2)}+\varepsilon_{1}\varepsilon_{2}\nabla_{i}\right]\nonumber \\
 &   +\sum_{1\leq i_{1}<i_{2}\leq r}\frac{z_{i_{1}}z_{i_{2}}}{\left(z_{0}-z_{i_{1}}\right)\left(z_{0}-z_{i_{2}}\right)}\left(\mathcal{A}_{i_{1}}^{(1)}-\varepsilon\right)\mathcal{A}_{i_{2}}^{(1)}\Bigg\}\mathcal{Z}^{\mathrm{instanton}}. \label{eq:bpz2gauge}
\end{align}
This is the equation that was derived in \cite{Nekrasov:2017gzb} to confirm the BPS/CFT correspondence for this particular case.

\subsubsection{Third-order differential equation}

The derivation can be extended to the next-to-simplest case, as we now explain. To obtain a third-order differential equation, we tune the parameters
\begin{equation}
a_{0,1}=a_{1,1}+2\varepsilon_{1},\quad a_{0,2}=a_{1,2}.
\end{equation}
In this case, the configurations of the gauge field are required to
satisfy that the Young diagram $Y^{(1,1)}$ has at most two rows and
$Y^{(1,2)}=\emptyset$,
\begin{equation}
Y^{(1)}=\left(\ytableausetup{mathmode,boxsize=3em, centertableaux}
\begin{ytableau}(1,1)&(1,2)&\none[\dots]&(1,y_{2})&\none[\dots]&(1,y_{1})\\
(2,1)&(2,2)&\none[\dots]&(2,y_{2})
\end{ytableau},\emptyset\right),
\end{equation}
where we denote the number of boxes in the first and the second row
of the Young diagram $Y^{(1,1)}$ as $y_{1}$ and $y_{2}$, respectively.
The instanton charge $k_{1}=y_{1}+y_{2}$. Then, we have
\begin{align}
\Xi_{0}(x)[\boldsymbol{Y}] & =  \frac{\mathscr{Y}_{1}(x+\varepsilon)[\boldsymbol{Y}]}{\mathscr{Y}_{0}(x)}\nonumber \\
 & =  \frac{\left(x+\varepsilon-a_{0,1}\right)\left(x+\varepsilon-a_{0,2}\right)}{\left(x-a_{0,1}\right)\left(x-a_{0,2}\right)}\nonumber \\
 &   \times\frac{x+\varepsilon-a_{0,1}+2\varepsilon_{1}-\varepsilon_{2}y_{1}}{x+\varepsilon-a_{0,1}+\varepsilon_{1}-\varepsilon_{2}y_{1}}\frac{x+\varepsilon-a_{0,1}+\varepsilon_{1}-\varepsilon_{2}y_{2}}{x+\varepsilon-a_{0,1}-\varepsilon_{2}y_{2}}\nonumber \\
 & = 1+\sum_{n=1}^{\infty}\frac{\zeta_{0,n}[\boldsymbol{Y}]}{x^{n}}.
\end{align}
We have 
\begin{align}
\zeta_{0,1} & =  4\varepsilon_{1}+2\varepsilon_{2},\nonumber \\
\zeta_{0,2} & =  \left(3\varepsilon_{1}+\varepsilon_{2}\right)a_{0,1}+\varepsilon a_{0,2}+\varepsilon\left(3\varepsilon_{1}+\varepsilon_{2}\right)+\varepsilon_{1}\varepsilon_{2}k_{1},
\end{align}
while $\zeta_{0,4}$ are related to $\zeta_{0,3}$ as 
\begin{align}
\zeta_{0,4} & =  \left(3a_{0,1}-2\varepsilon_{1}-\varepsilon_{2}+\frac{3}{2}\varepsilon_{2}k_{1}\right)\zeta_{0,3}-\frac{1}{2}\varepsilon_{1}\varepsilon_{2}^{3}k_{1}^{3}-\frac{1}{2}\varepsilon_{1}\varepsilon_{2}^{2}\left(6a_{0,1}-\varepsilon_{1}\right)k_{1}^{2}\nonumber \\
 &   -\varepsilon_{2}\left(\frac{3}{2}\left(5\varepsilon_{1}+\varepsilon_{2}\right)a_{0,1}^{2}+\frac{1}{2}\left(\varepsilon_{1}+3\varepsilon_{2}\right)\left(3\varepsilon_{1}+\varepsilon_{2}\right)a_{0,1}+\frac{3}{2}\varepsilon a_{0,2}^{2}+\frac{1}{2}\varepsilon\left(7\varepsilon_{1}+3\varepsilon_{2}\right)a_{0,2}\right)k_{1}\nonumber \\
 &   -2\left(3\varepsilon_{1}+\varepsilon_{2}\right)a_{0,1}^{3}-\varepsilon_{2}\left(3\varepsilon_{1}+\varepsilon_{2}\right)a_{0,1}^{2}+\varepsilon\left(2\varepsilon_{1}+\varepsilon_{2}\right)\left(3\varepsilon_{1}+\varepsilon_{2}\right)a_{0,1}-3\varepsilon a_{0,1}a_{0,2}^{2}\nonumber \\
 &   -2\varepsilon\left(3\varepsilon_{1}+\varepsilon_{2}\right)a_{0,1}a_{0,2}+\varepsilon a_{0,2}^{3}+\varepsilon\left(5\varepsilon_{1}+2\varepsilon_{2}\right)a_{0,2}+\varepsilon\left(2\varepsilon_{1}+\varepsilon_{2}\right)\left(3\varepsilon_{1}+\varepsilon_{2}\right)a_{0,2}.\label{eq:zeta04}
\end{align}
Using (\ref{eq:NPDS-10}), (\ref{eq:zeta04}), and (\ref{eq:NPDS-1j}),
we can get rid of all terms with $\langle\zeta_{0,3}\rangle$, $\langle\zeta_{0,4}\rangle$,
and $\langle\zeta_{i,3}\rangle$ in (\ref{eq:NPDS-20}), and we obtain
a differential equation on the instanton partition function
\begin{align}
0 & =  \Bigg\{\frac{\varepsilon_{2}^{3}}{2}\nabla_{0}^{3}+\frac{\varepsilon_{2}^{2}}{2}\left(-3a_{0,1}+3a_{0,2}+\varepsilon_{1}+3\sum_{i=1}^{r}\frac{z_{i}}{z_{0}-z_{i}}\mathcal{A}_{i}^{(1)}\right)\nabla_{0}^{2}\nonumber \\
 &   +\varepsilon_{2}\Bigg[\left(a_{0,1}-a_{0,2}\right)\left(a_{0,1}-a_{0,2}-\varepsilon_{1}\right)\nonumber \\
 &   -\sum_{i=1}^{r}\frac{z_{i}}{z_{0}-z_{i}}\left(2\mathcal{A}_{i}^{(2)}-2\varepsilon_{1}\varepsilon_{2}\nabla_{i}+\mathcal{A}_{i}^{(1)}\left(2a_{0,1}-4a_{0,2}+\frac{\varepsilon_{2}}{2}\right)\right)\nonumber \\
 &   +\sum_{i=1}^{r}\frac{z_{i}^{2}}{\left(z_{0}-z_{i}\right)^{2}}\mathcal{A}_{i}^{(1)}\left(\mathcal{A}_{i}^{(1)}-\varepsilon_{1}-\frac{\varepsilon_{2}}{2}\right)\nonumber \\
 &   +2\sum_{1\leq i_{1}<i_{2}\leq r}\frac{z_{i_{1}}z_{i_{2}}}{\left(z_{0}-z_{i_{1}}\right)\left(z_{0}-z_{i_{2}}\right)}\left(2\mathcal{A}_{i_{1}}^{(1)}-\varepsilon\right)\mathcal{A}_{i_{2}}^{(1)}\Bigg]\nabla_{0}\nonumber \\
 &  -\sum_{i=1}^{r}\frac{z_{i}}{z_{0}-z_{i}}\left(2a_{0,1}-2a_{0,2}+\varepsilon_{2}\right)\left(a_{0,2}\mathcal{A}_{i}^{(1)}-\mathcal{A}_{i}^{(2)}+\varepsilon_{1}\varepsilon_{2}\nabla_{i}\right)\nonumber \\
 &   +\sum_{i=1}^{r}\frac{z_{i}^{2}}{\left(z_{0}-z_{i}\right)^{2}}\left(2\mathcal{A}_{i}^{(1)}-2\varepsilon_{1}-\varepsilon_{2}\right)\left(a_{0,2}\mathcal{A}_{i}^{(1)}-\mathcal{A}_{i}^{(2)}+\varepsilon_{1}\varepsilon_{2}\nabla_{i}\right)\nonumber \\
 & +2\sum_{1\leq i_{1}<i_{2}\leq r}\frac{z_{i_{1}}z_{i_{2}}}{\left(z_{0}-z_{i_{1}}\right)\left(z_{0}-z_{i_{2}}\right)}\Bigg[\left(\mathcal{A}_{i_{1}}^{(1)}-\varepsilon\right)\mathcal{A}_{i_{2}}^{(1)}\left(-a_{0,1}+a_{0,2}-\varepsilon\right)\nonumber \\
 &   +\mathcal{A}_{i_{1}}^{(1)}\left(-\mathcal{A}_{i_{2}}^{(2)}+a_{0,2}\mathcal{A}_{i_{2}}^{(1)}+\varepsilon_{1}\varepsilon_{2}\nabla_{i}\right)\nonumber \\
 &   +\mathcal{A}_{i_{2}}^{(1)}\left(-\mathcal{A}_{i_{1}}^{(2)}+a_{0,2}\mathcal{A}_{i_{1}}^{(1)}+\varepsilon_{1}\varepsilon_{2}\nabla_{i}\right)\Bigg]\nonumber \\
 &   +2\sum_{1\leq i_{1}<i_{2}<i_{3}\leq r}\frac{z_{i_{1}}z_{i_{2}}z_{i_{3}}}{\left(z_{0}-z_{i_{1}}\right)\left(z_{0}-z_{i_{2}}\right)\left(z_{0}-z_{i_{3}}\right)}\left(\mathcal{A}_{i_{1}}^{(1)}\left(3\mathcal{A}_{i_{2}}^{(1)}-\varepsilon\right)\mathcal{A}_{i_{3}}^{(1)}-2\varepsilon\mathcal{A}_{i_{2}}^{(1)}\mathcal{A}_{i_{3}}^{(1)}\right)\nonumber \\
 &   +\sum_{1\leq i_{1}<i_{2}\leq r}\frac{z_{i_{1}}^{2}z_{i_{2}}}{\left(z_{0}-z_{i_{1}}\right)^{2}\left(z_{0}-z_{i_{2}}\right)}\left(\mathcal{A}_{i_{1}}^{(1)}-\varepsilon\right)\left(2\mathcal{A}_{i_{1}}^{(1)}-2\varepsilon_{1}-\varepsilon_{2}\right)\mathcal{A}_{i_{2}}^{(1)}\nonumber \\
 &   +\sum_{1\leq i_{1}<i_{2}\leq r}\frac{z_{i_{1}}z_{i_{2}}^{2}}{\left(z_{0}-z_{i_{1}}\right)\left(z_{0}-z_{i_{2}}\right)^{2}}\left(\mathcal{A}_{i_{1}}^{(1)}-\varepsilon\right)\mathcal{A}_{i_{2}}^{(1)}\left(2\mathcal{A}_{i_{2}}^{(1)}-2\varepsilon_{1}-\varepsilon_{2}\right)\Bigg\}\mathcal{Z}^{\mathrm{instanton}}. \label{eq:bpz3gauge}
\end{align}

\subsection{Identification with the BPZ equations}

The final step is to identify the differential equations we derived
in the gauge theory side with the BPZ equations by solving the undetermined
parameters $L_{i}$ and $T_{ij}$. We find that the following
solution exists:
\begin{align}
L_{0} & =  \Delta_{-1}-\Delta_{0}-\frac{\varepsilon^{2}-(a_{1,1}-a_{1,2})^{2}}{4\varepsilon_{1}\varepsilon_{2}},\nonumber \\
L_{i} & =  -\Delta_{i}-\frac{\left(a_{i,1}-a_{i,2}\right)^{2}}{4\varepsilon_{1}\varepsilon_{2}}+\frac{\left(a_{i+1,1}-a_{i+1,2}\right)^{2}}{4\varepsilon_{1}\varepsilon_{2}},\quad i=1,\cdots,r-1,\nonumber \\
L_{r} & =  -\Delta_{r}-\Delta_{r+1}+\frac{\varepsilon^{2}-\left(a_{r,1}-a_{r,2}\right)^{2}}{4\varepsilon_{1}\varepsilon_{2}},\nonumber \\
T_{ij} & =  \frac{\left(\mathcal{A}_{i}^{(1)}-2\varepsilon\right)\mathcal{A}_{j}^{(1)}}{2\varepsilon_{1}\varepsilon_{2}}. \label{eq:identify}
\end{align}
With this identification of parameters, we observe the precise agreement between (\ref{eq:bpz2}), (\ref{eq:bpz3}), and (\ref{eq:bpz2gauge}), (\ref{eq:bpz3gauge}). It is easy to check that (\ref{eq:Ward2-3}) is also satisfied. Notice
that the prefactor can also be written as
\begin{align}
\left(\prod_{i=0}^{r}z_{i}^{L_{i}}\right)\prod_{0\leq i<j\leq r}\left(1-\frac{z_{j}}{z_{i}}\right)^{T_{ij}} & =  \left(z_{0}^{\Delta_{-1}-\Delta_{0}-\frac{\varepsilon^{2}}{4\varepsilon_{1}\varepsilon_{2}}}\right)\left(\prod_{i=1}^{r-1}z_{i}^{-\Delta_{i}}\right)\left(z_{r}^{\frac{\varepsilon^{2}}{4\varepsilon_{1}\varepsilon_{2}}-\Delta_{r}-\Delta_{r+1}}\right)\nonumber \\
 &   \times\prod_{i=1}^{r}q_{i}^{-\frac{\left(a_{i,1}-a_{i,2}\right)^{2}}{4\varepsilon_{1}\varepsilon_{2}}}\prod_{0\leq i<j\leq r}\left(1-\frac{z_{j}}{z_{i}}\right)^{\frac{2\left(\bar{a}_{i}-\bar{a}_{i+1}\right)\left(\bar{a}_{j}-\bar{a}_{j+1}+\varepsilon\right)}{\varepsilon_{1}\varepsilon_{2}}}.
\end{align}
which give the expected tree-level partition function and the $\mathrm{U}(1)$
part of the partition function (see Appendix \ref{sec:U1} for details). Therefore, we confirm the BPS/CFT correspondence.

\section{Conclusions and future directions \label{sec:Conclusions}}

In this paper, we perform the derivation of the differential equation on the instanton partition function at a special point in the parameter space using the method of NPDS equations and identify the differential equations with the BPZ equations in the Liouville field theory. Therefore, we confirm the main assertion of the BPS/CFT correspondence. It is interesting to notice that this application of NPDS equations is complementary to the study of chiral trace relations in \cite{Jeong:2019fgx}, where parameters are taken to be generic.

There are several obvious generalizations of our paper. First of all, it is natural to consider the general degenerate fields with conformal dimension $\Delta_{(m,n)}$ and derive the differential equation of order $mn$ on the instanton partition function. The computation will be unavoidably lengthy, but the basic idea is the same. To simplify the derivation, it is sometimes useful to consider the NPDS equations of both fundamental and nonfundamental qq-characters.

We can also generalize the discussion to the $\mathrm{U}(N)$ superconformal linear quiver gauge theories. However, from the knowledge of corresponding Toda field theory, we do not expect to obtain a differential equation on the instanton partition function. Instead, the equations derived from the NPDS equations will generally relate the instanton partition function with the expectation values of certain BPS observables. Only if we take the Nekrasov-Shatashvili limit can we get a differential equation on the instanton partition function. The nonconformal $A_2$-quiver $\mathrm{SU}(3)$ gauge theory and the degenerate irregular conformal block in the $A_2$ Toda field theory were studied in \cite{Jeong:2017pai} along this direction. The detailed discussion on the general quiver will appear in a separate paper.

In spite of the successful application of the NPDS equations to derive the BPZ equations, there are still some open problems. From the point of view of conformal field theory, it is equally good to choose any one of the fields to be degenerate, and we have the BPZ equation for every choice. In the corresponding four-dimensional theory, we need to tune the parameters in the following way for arbitrary $i=0,\cdots,r$:
\begin{equation}
a_{i,\alpha}=\begin{cases}
a_{i+1,1}+(m-1)\varepsilon_{1}+(n-1)\varepsilon_{2}, & \alpha=1,\\
a_{i+1,\alpha}, & \alpha\neq 1.
\end{cases}\label{eq:masstuning2}
\end{equation}
However, we do not get the expected constraints of the form (\ref{eq:masstuning2}). For example, the constraint is $Y^{(i+1,\alpha)}\subset Y^{(i,\alpha)}$ rather than $Y^{(i+1,\alpha)}=Y^{(i,\alpha)}$ for $\alpha\neq1$. This problem is associated with the annoying $\mathrm{U}(1)$ factor in the AGT dictionary. We may have to figure out how to factor out the $\mathrm{U}(1)$ factor at the level of the measure $\mathcal{Z}^{\mathrm{instanton}}\left(\boldsymbol{a};\boldsymbol{Y};\varepsilon_{1},\varepsilon_{2}\right)$. Progress in this direction will also lead us immediately to a derivation of the BPZ equation for the conformal field theory on a torus.

\begin{acknowledgments}
The authors are deeply indebted to Nikita Nekrasov for invaluable suggestions and discussions. The authors also would like to thank Alex DiRe, Wolfger Peelaers, and Naveen Prabhakar for helpful discussions. The work was supported in part by the National Science Foundation under Grant No. NSF-PHY 1404446. The work of S. J.  was also supported by the Samsung Scholarship. X. Z. was also funded by National Science Foundation of China under Grant No. 12347103, and the DOE under Grant No. DOE-SC0010008.
\end{acknowledgments}

\appendix

\section{Generalized hypergeometric function \label{sec:hypergeometric}}

A generalized hypergeometric function $_{p}F_{q}\left(a_{1},\cdots,a_{p};b_{1},\cdots,b_{q};z\right)$
is defined as the power series
\begin{equation}
_{p}F_{q}\left(a_{1},\cdots,a_{p};b_{1},\cdots,b_{q};z\right)=\sum_{n=0}^{\infty}\frac{a_{1}^{\bar{n}}\cdots a_{p}^{\bar{n}}}{b_{1}^{\bar{n}}\cdots b_{q}^{\bar{n}}}\frac{z^{n}}{n!},\label{eq:pFqDef}
\end{equation}
where the parameters $a_{1},\cdots,a_{p},b_{1},\cdots,b_{q}$ are
complex numbers. The notation $x^{\bar{n}}$ is the rising factorial
or Pochhammer function, which is defined for real values of $n$ using
the Gamma function provided that $x$ and $x+n$ are real numbers
that are not negative integers,
\begin{equation}
x^{\bar{n}}=\frac{\Gamma(x+n)}{\Gamma(x)}.
\end{equation}
The generalized hypergeometric function $_{p}F_{q}\left(a_{1},\cdots,a_{p};b_{1},\cdots,b_{q};z\right)$
is a solution to the generalized hypergeometric function,
\begin{equation}
\left[z\prod_{n=1}^{p}\left(z\frac{d}{dz}+a_{n}\right)-z\frac{d}{dz}\prod_{n=1}^{q}\left(z\frac{d}{dz}+b_{n}-1\right)\right]{}_{p}F_{q}\left(a_{1},\cdots,a_{p};b_{1},\cdots,b_{q};z\right)=0.
\end{equation}
Clearly, the order of the parameters $\left\{ a_{1},\cdots,a_{p}\right\} $,
or the order of the parameters $\left\{ b_{1},\cdots,b_{q}\right\} $
can be changed without changing the value of the function. The standard
hypergeometric function $_{2}F_{1}\left(a,b;c;z\right)$ is simply
a special case of the generalized hypergeometric function when $p=2$
and $q=1$.

If any $a_{j}$ is a nonpositive integer, then the series (\ref{eq:pFqDef})
only has a finite number of terms and becomes a polynomial of degree
$-a_{j}$. If any $b_{k}$ is a nonpositive integer (excepting the
previous case with $b_{k}<a_{j}$), then the series (\ref{eq:pFqDef})
is undefined. Excluding these special cases for which the numerator
or the denominator of the coefficients can be $0$, the radius of
convergence can be determined using the ratio test. In this paper,
we are interested in the case $p=q+1$. The ratio of coefficients
tends to one, implying that the series (\ref{eq:pFqDef}) converges
for $|z|<1$ and diverges for $|z|>1$.

\section{$\mathrm{U}(1)$ factor \label{sec:U1}}

In this appendix, we derive the $\mathrm{U}(1)$ factor using the NPDS equations.

When $N=1$, we have $\mathscr{Y}_{0}(x)=x-\bar{a}_{0}$, and the
 NPDS equations lead to
\begin{align}
0 & =  \langle\left[\mathscr{Y}_{0}(x)\mathscr{G}_{r}(x;t)\right]^{(-1)}\rangle\rangle\nonumber \\
 & =  \langle\mathscr{G}_{r}^{(-2)}(t)\rangle-\bar{a}_{0}\langle\mathscr{G}_{r}^{(-1)}(t)\rangle\nonumber \\
 & =  \langle U_{r}[2]\rangle+\langle U_{r}[1,1]\rangle-\varepsilon\langle U_{r}[0,1]\rangle-\bar{a}_{0}\langle U_{r}[1]\rangle\nonumber \\
 & =  \sum_{i=0}^{r}u_{i}\left(\langle\zeta_{i,2}\rangle-\bar{a}_{0}\langle\zeta_{i,1}\rangle\right)+\sum_{0\leq i_{1}<i_{2}\leq r}u_{i_{1}}u_{i_{2}}\left(\langle\zeta_{i_{1},1}\zeta_{i_{2},1}\rangle-\varepsilon\langle\zeta_{i_{2},1}\rangle\right),
\end{align}
where 
\begin{align}
\zeta_{j,1} & =  \bar{a}_{j}-\bar{a}_{j+1}+\varepsilon,\\
\zeta_{j,2} & =  \bar{a}_{j}\left(\bar{a}_{j}-\bar{a}_{j+1}+\varepsilon\right)-\varepsilon_{1}\varepsilon_{2}\left(k_{i}-k_{i+1}\right).
\end{align}
Picking up the residue at $t=-z_{j}^{-1}$ we obtain,
\begin{align}
\varepsilon_{1}\varepsilon_{2}\langle k_{j}-k_{j+1}\rangle & =  \left(\bar{a}_{j}-\bar{a}_{0}\right)\left(\bar{a}_{j}-\bar{a}_{j+1}+\varepsilon\right)\nonumber \\
 &   +\sum_{i=j+1}^{r}\frac{z_{i}}{z_{i}-z_{j}}\left(\bar{a}_{j}-\bar{a}_{j+1}\right)\left(\bar{a}_{i}-\bar{a}_{i+1}+\varepsilon\right)\nonumber \\
 &   +\sum_{i=0}^{j-1}\frac{z_{i}}{z_{i}-z_{j}}\left(\bar{a}_{i}-\bar{a}_{i+1}\right)\left(\bar{a}_{j}-\bar{a}_{j+1}+\varepsilon\right)\nonumber \\
 & =  \sum_{i=j+1}^{r}\frac{z_{i}}{z_{i}-z_{j}}\left(\bar{a}_{j}-\bar{a}_{j+1}\right)\left(\bar{a}_{i}-\bar{a}_{i+1}+\varepsilon\right)\nonumber \\
 &   +\sum_{i=0}^{j-1}\frac{z_{j}}{z_{i}-z_{j}}\left(\bar{a}_{i}-\bar{a}_{i+1}\right)\left(\bar{a}_{j}-\bar{a}_{j+1}+\varepsilon\right).
\end{align}
From the structure of the instanton partition function, we see that
\begin{equation}
\langle k_{j}-k_{j+1}\rangle=z_{j}\frac{d}{dz_{j}}\log Z^{\mathrm{instanton}}.
\end{equation}
Therefore, we have the $\mathrm{U}(1)$ part of the instanton partition function
\begin{equation}
Z^{\mathrm{instanton}}=\prod_{0\leq i<j\leq r}\left(1-\frac{z_{j}}{z_{i}}\right)^{-\frac{\left(\bar{a}_{i}-\bar{a}_{i+1}\right)\left(\bar{a}_{j}-\bar{a}_{j+1}+\varepsilon\right)}{\varepsilon_{1}\varepsilon_{2}}}.
\end{equation}

\bibliography{BPSNDS-Ref}

\end{document}